\documentclass[11pt,letterpaper]{article}

\usepackage{typearea}
\typearea{15}

\usepackage{theorem,latexsym,graphicx}
\usepackage{amsmath,amssymb,enumerate}
\usepackage{xspace}
\usepackage{bm}
\usepackage{ifpdf}
\usepackage{shadow,shadethm,color}
\usepackage[compact]{titlesec}
\usepackage{algorithm, algorithmic}
\usepackage{array,tabularx,mdwtab}


\definecolor{Darkblue}{rgb}{0,0,0.4}
\definecolor{Brown}{cmyk}{0,0.81,1.,0.60}
\definecolor{Purple}{cmyk}{0.45,0.86,0,0}

\ifpdf
 
\fi
\newcommand{\lref}[2][]{\hyperref[#2]{#1~\ref*{#2}}}
\newcommand{\vect}[1]{\boldsymbol{#1}}

\newtheorem{theorem}{Theorem}[section]
\newtheorem{definition}[theorem]{Definition}

\newtheorem{lemma}[theorem]{Lemma}
\newshadetheorem{lemmashaded}[theorem]{Lemma}
\newtheorem{fact}[theorem]{Fact}
\newtheorem{claim}[theorem]{Claim}
\newtheorem{myclaim}[theorem]{Claim}
\newtheorem{example}[theorem]{Example}

\newtheorem{corollary}[theorem]{Corollary}

\newenvironment{proof}{

\noindent{\bf Proof:}}
{\hfill$\blacksquare$

}

\newenvironment{proofof}[1]{

\noindent{\bf Proof of {#1}:}}
{\hfill$\blacksquare$

}

\newcommand{\initOneLiners}{%
    \setlength{\itemsep}{0pt}
    \setlength{\parsep }{0pt}
    \setlength{\topsep }{0pt}
}
\newenvironment{OneLiners}[1][\ensuremath{\bullet}]
    {\begin{list}
        {#1}
        {\initOneLiners}}
    {\end{list}}

\def\abs#1{\mathopen| #1 \mathclose|}	
\def\norm#1{\mathopen\| #1 \mathclose\|}
\def\qed{}

\begin{document}

\title{A Direct Reduction from $k$-Player to $2$-Player Approximate Nash
Equilibrium\thanks{Work supported in part by The Israel Science Foundation (grant No. 873/08).}}

\author{
  Uriel Feige
\thanks{Weizmann Institute of Science, PO Box 26, Rehovot 76100,
Israel. Email: \texttt{uriel.feige@weizmann.ac.il}. The author holds the Lawrence G. Horowitz Professorial Chair at the Weizmann Institute.}
  \\Weizmann Institute
  \and
  Inbal Talgam
\thanks{Weizmann Institute of Science, PO Box 26, Rehovot 76100,
Israel. Email: \texttt{inbal.talgam@weizmann.ac.il}.}
  \\Weizmann Institute
}

  \date{}
  \maketitle

  \begin{abstract}\noindent
We present a direct reduction from $k$-player games to 2-player games
that preserves approximate Nash equilibrium. Previously, the computational
equivalence of computing approximate Nash equilibrium in $k$-player
and 2-player games was established via an indirect reduction. This
included a sequence of works defining the complexity class PPAD, identifying
complete problems for this class, showing that computing approximate
Nash equilibrium for $k$-player games is in PPAD, and reducing a PPAD-complete
problem to computing approximate Nash equilibrium for 2-player games.
Our direct reduction makes no use of the concept of PPAD, thus eliminating some of the difficulties involved in following the known indirect reduction. 
  \end{abstract}

\section{Introduction}
\label{sec:Introduction}

This manuscript addresses the computation of Nash equilibrium for games
represented in normal form. It is known that for 2-player games this
problem is PPAD-complete \cite{CD06a}, and for $k$ players it is in PSPACE \cite{EY07}. Moreover, for sufficiently small $\epsilon$, computing $\epsilon$-well-supported Nash equilibrium for 2-player games remains PPAD-complete \cite{CDT06}, and for $k$ players it is in PPAD \cite{DGP09}. It follows that, for appropriate choices of $\epsilon$, $\epsilon$-well-supported
Nash in $k$-player games reduces to $\epsilon$-well-supported
Nash in 2-player games. However, this chain of reductions is indirect,
passing through intermediate notions other than games, and also rather complicated.

In this manuscript we present a direct, "game theoretic" polynomial-time reduction from $k$-player to 2-player games. In our reduction, every pure strategy of each of the $k$ players is represented by a corresponding pure strategy of one of the 2 players. Previously, a direct reduction preserving exact Nash equilibrium was known from $k$-player to 3-player games \cite{Bub79}. Such a reduction cannot exist to 2-player games due to issues of irrationality \cite{Nas51}, hence the need to consider the notion of $\epsilon$-well-supported Nash in this context. Our reduction guarantees that for appropriate choices of $\epsilon_2$ and $\epsilon_k$, given any $\epsilon_2$-well-supported Nash for the 2-player game, normalizing its probabilities according to the above correspondence immediately gives an $\epsilon_k$-well-supported Nash for the $k$-player game.

The direct reduction makes no use of the concept of PPAD. This eliminates
some of the difficulties involved in following the known indirect
reduction. It is inevitable that unlike the indirect reduction, our
reduction by itself does not establish the PPAD-completeness of computing
(or approximating) Nash equilibria. Nevertheless, the new gadgets we introduce are relevant to the notion of PPAD-completeness, as they can be used in other reductions among PPAD problems. Moreover, our reduction provides an alternative proof to the proof of Daskalakis et al.~\cite{DGP09} that finding an approximate Nash equilibrium in $k$-player games is in PPAD. 

In the $k$-player case, the payoff of each player depends on the combined behavior of the other players. We can thus view each player's set of expected payoffs as a set of multiplicative functions in the other players' strategies. In a 2-player game, however, each player interacts with a single other player, and so the expected payoffs are linear \cite{vSte07}. The described gap calls for "linearization" of $k$-player games, and indeed the first step of our reduction replaces the multilateral interactions among the $k$ players with bilateral interactions among pairs of players. In the next step, two representative "super-players" replace the multiple players, resulting in a 2-player game. 

In terms of techniques, the first step of the reduction uses and extends the machinery of gadget games developed by Goldberg and Papadimitriou \cite{GP06}. We introduce a new gadget for performing approximate multiplication using linear operations, in order to bridge the gap between multiplicative and linear games. The second step of the reduction uses similar methods to \cite{GP06} and \cite{MT09} in order to replace multiple players by 2 players. The resulting 2-player game is a combination of a generalized Matching Pennies game \cite{GP06} and an imitation game \cite{MT09}.

\subsection{Preliminaries}

Let $[n]=\{1,\dots,n\}$, and $\norm{v}=\sum_i{\abs{v_i}}$. For vectors $\vect{u}$ and $\vect{v}$ of length $n$, let $\vect{u}\otimes\vect{v}$ denote their tensor product written as a vector of length $n^2$, where entry $(i-1)n+j$ is $u_iv_j$. We write $x=y\pm z$ to denote $y-z\le x\le y+z$. For vectors, $\vect{x}=\vect{y}\pm z$ denotes $y_i-z\le x_i\le y_i+z$ for every $i$. 

\paragraph{Normal Form Games}

Players of a normal form game $G_k$ are numbered from 1 to $k$. Unless stated otherwise, every player has $n$ pure strategies numbered from 1 to $n$. A \emph{pure strategy profile} $\vect{s}$ is a vector of length $k$ in $[n]\times\dots\times[n]$, containing one pure strategy per player. $\vect{s^{-i}}$ is a pure strategy profile for all players except $i$, obtained from $\vect{s}$ by removing the $i$'th entry. A \emph{payoff matrix} $M^i=M^i_{G_k}$ for player $i$ is of size $n\times n^{k-1}$. Unless stated otherwise, all entries are rationals in the $[0,1]$ range. $M^i[j,\vect{s^{-i}}]$ is the payoff player $i$ receives for playing pure strategy $j$ against pure strategy profile $\vect{s^{-i}}$. A \emph{mixed strategy} $\vect{p^i}$ for player $i$ is a probability distribution over $[n]$, denoting the probabilities with which $i$ plays her pure strategies. Its \emph{support} is the set of pure strategies $\{j:p^i_j>0\}$. A \emph{mixed strategy profile} $\vect{p}=(\vect{p^1},\dots,\vect{p^k})$
is a set of mixed strategies for every player, and $\vect{p^{-i}}$ is a similar set for every player except $i$. Let $\vect{\tilde{p}}$ be the \emph{joint mixed strategy} distribution, i.e. $\vect{\tilde{p}}=\vect{p^1}\otimes\dots\otimes\vect{p^k}$. For every pure strategy profile $\vect{s}$, entry $\vect{\tilde{p}}[\vect{s}]$ is the probability $\prod_i{p^i_{s_i}}$ that every player $i$ plays pure strategy $s_i$. Let $\vect{\tilde{p}^{-i}}$ be the joint mixed strategy of all players except $i$. Given a mixed strategy profile $\vect{p^{-i}}$, the \emph{expected payoff vector} $\vect{u^i_{G_k}}$ equals $M^i\vect{\tilde{p}^{-i}}$. The $j$'th entry $\vect{u^i_{G_k}}[j]$ is the expected payoff player $i$ receives for playing pure strategy $j$ while the others play $\vect{p^{-i}}$. Thus, the expected payoffs are algebraic functions in the probabilities played by the others. 

\paragraph{Polymatrix (Linear) Games}

In a polymatrix game, every player plays bilaterally against other players, and receives the sum of payoffs obtained from these bilateral interactions. Thus, polymatrix games are actually collections of 2-player games in which every player plays the same strategy in every game she participates in. Players are numbered from 1 to $m$; player $i$ has $2\le n_i\le n$ pure strategies and $m-1$ rational payoff matrices $M^{i,i'}$ of size $n_i\times n_{i'}$. Entry $M^{i,i'}[j,j']$ is the payoff to player $i$ for playing $j$ against player $i'$ who plays $j'$. If players $i,i'$ do not interact or if their interaction is one-sided and does not influence player $i$'s payoff, then $M^{i,i'}$ is set to be all-zeros. Given a pure strategy profile $\vect{s^{-i}}$, the total payoff to player $i$ for playing $j$ is $\sum_{i'\ne i}{M^{i,i'}[j,\vect{s^{-i}}[i']]}$. Given a mixed strategy profile $\vect{p^{-i}}$, the expected payoff vector of player $i$ is $\vect{u^i_{G_m}}=\sum_{i'\ne i}{M^{i,i'}}\vect{p^{i'}}$. Equivalently, if $M^i_{G_m}=(M^{i,1}\cdots M^{i,m})$ contains all the player's payoff matrices as submatrices, then $\vect{u^i_{G_m}}=M^i_{G_m}\vect{p^{-i}}$. The expected payoffs of a player are thus linear functions in the probabilities of the others. Unlike normal form games, the size of polymatrix games is polynomial in $n$ even when the number of players $m$ is non-constant ($m=\mbox{poly}(n)$).

\paragraph{Nash Equilibrium, Approximations and Computational Problems}

A \emph{Nash equilibrium} is a mixed strategy profile such that the players of the game cannot improve their expected payoffs by deviating from it unilaterally. The supports of a Nash equilibrium contain only pure strategies that are \emph{best responses}, i.e. maximize the expected payoff given the mixed strategies of the other players. Formally, given a mixed strategy profile $\vect{p^{-i}}$, pure strategy $j$ is a best response for player $i$ if $\vect{u_G^i}[j]=\max_{j'\in[n]}\{\vect{u_G^i}[j']\} $.

Every game has a Nash equilibrium \cite{Nas51}, but finding such an equilibrium may be difficult. There are games for which every Nash equilibrium contains irrational probabilities, making it hard even to represent. This motivates the consideration of approximate instead of exact Nash equilibrium. In the context of reductions from $k$-player to $2$-player games, there is another motivation for considering approximate Nash equilibrium. Unlike $k$ player games, $2$-player games always have a rational Nash equilibrium \cite{Nas51,Pap07}. Thus we do not expect to find a reduction that preserves exact Nash equilibrium, direct or indirect.

There are several possible notions of approximation. We focus on the notion of \emph{$\epsilon$-well-supported} Nash equilibrium, a mixed strategy profile whose supports contain only \emph{$\epsilon$-}best responses, i.e. pure strategies that maximize the expected payoff up to an additive factor of $\epsilon$. We will primarily be interested in small, non-constant values of $\epsilon$, namely $\epsilon=1/\mbox{poly}(n)$ and $\epsilon=1/\exp(n)$. A related, computationally equivalent approximation notion is that of $\epsilon$-Nash equilibrium - a mixed strategy profile from which deviating unilaterally cannot improve a player's expected payoff by more than $\epsilon$ \cite{DGP09}. For other approximation notions see \cite{EY07}.

\begin{definition}[$\epsilon_k$-$k$NASH and $\epsilon_m$-LINEAR-NASH]
Given a pair of normal form game $G_k$ and accuracy parameter $\epsilon_k$, the problem $\epsilon_k$-$k$NASH is to find an $\epsilon_k$-well-supported Nash equilibrium of $G_k$. Given a pair of polymatrix game $G_m$ and accuracy parameter $\epsilon_m$, the problem $\epsilon_m$-LINEAR-NASH is to find an $\epsilon_m$-well-supported Nash equilibrium of $G_m$.
\end{definition}

\subsection{Our Results}

Let $(G_{m_1},\epsilon_{m_1}),(G_{m_2},\epsilon_{m_2})$ be two pairs of games and accuracy parameters. The games have $m_1,m_2$ players respectively; the number of pure strategies of player $i$ is $n^1_i,n^2_i$ respectively. The following definitions are based on the notion of \emph{reduction scheme} defined by Bubelis \cite{Bub79}.

\begin{definition}[Mapping between Games]
\label{def:mapping}
A \emph{mapping} from $G_{m_1}$ to $G_{m_2}$ includes:
\begin{OneLiners}
\item A function $g:[m_1]\rightarrow[m_2]$ mapping players of $G_{m_1}$ to players  of $G_{m_2}$;
\item For every $i\in[m_1]$, an injective function $h_i:[n^1_i]\rightarrow[n^2_{g(i)}]$ mapping pure strategies of player $i$ to distinct pure strategies of player $g(i)$. 
\end{OneLiners}
\end{definition}

\begin{definition}[Direct Reduction]
\label{def:direct-reduction}
A \emph{direct reduction} from $(G_{m_1},\epsilon_{m_1})$ to $(G_{m_2},\epsilon_{m_2})$ is a mapping from $G_{m_1}$ to $G_{m_2}$, such that for every $\epsilon_{m_2}$-well-supported Nash equlibrium $(\vect{q^1},\dots,\vect{q^{m_2}})$ of $G_{m_2}$, an $\epsilon_{m_1}$-well-supported Nash equilibrium $(\vect{p^1},\dots,\vect{p^{m_1}})$ of $G_{m_1}$ can be obtained by renormalizing probabilities as follows:
$\vect{p^i}[j]=(1/z)\vect{q^{g(i)}}[h_i(j)]$ (where $z$ is a normalization factor).
\end{definition}

\begin{theorem}[Main]
\label{thm:Main-thm}
For every $\epsilon_k<1$, there exists a direct reduction from $\epsilon_k$-$k$NASH to $\epsilon_2$-$2$NASH, where $\epsilon_2=\mbox{poly}(\epsilon_k/\abs{G_k})$. The reduction runs in polynomial time in $\abs{G_k}$ and in $\log(1/\epsilon_k)$.
\end{theorem}

\begin{corollary}
\label{cor:Cor}
There is a direct, polynomial time reduction from $(1/\exp(n))$-$k$NASH to $(1/\exp(n))$-$2$NASH, and from $(1/\mbox{poly}(n))$-$k$NASH to $(1/\mbox{poly}(n))$-$2$NASH.
\end{corollary}

\begin{proofof}{\lref[Theorem]{thm:Main-thm}}
By combining \lref[Theorem]{thm:first-reduction} (linearizing reduction) with \lref[Theorem]{thm:part2} (reduction from linear to bimatrix games), and plugging in the parameters of \lref[Lemma]{lem:construction2} (logarithmic-sized linear multiplication gadget). 
\end{proofof}

For simplicitly of presentation we defer the proof of \lref[Lemma]{lem:construction2} to \lref[Section]{sec:linear-mult-2},
and first prove in \lref[Section]{sec:linear_mult} a slightly weaker version (\lref[Lemma]{lem:construction1} - polynomial-sized linear multiplication gadget), resulting in a reduction that runs in polynomial time in $1/\epsilon_k$ instead of $\log(1/\epsilon_k)$.

\subsection{Related Work}
\label{sub:Related-Work}

Bubelis \cite{Bub79} shows a direct reduction from $k$-player to 3-player games. This reduction relies heavily on the multiplicative nature of 3-player games. Examples of direct reductions involving 2-player games include \emph{symmetrization} \cite{GKT50}, and reduction to \emph{imitation games} \cite{MT09}. We use imitation games in Section \ref{sec:second-part-of-reduction}. 

\paragraph{PPAD-completeness results}

Papadimitriou introduced PPAD in 1991, motivated largely by the challenge of classifying the Nash equilibrium problem \cite{Pap94}. Formally, PPAD is the class of total search problems polynomial-time reducible to the abstract path-following problem END OF THE LINE. Another important PPAD-complete problem is 3D-BROUWER, a discrete version of finding Brouwer fixed-points in a 3-dimensional domain (the same problem in high-dimension is known as $n$D-BROUWER). The known results can be summarized by the two following chains of reductions, each forming an indirect reduction (according to \lref[Definition]{def:direct-reduction} of directness) from $k$-player games to 2-player games:

\begin{OneLiners}
\item $1/\exp(n)$-$k$NASH $\le$ END OF THE LINE $\le$ 3D-BROUWER $\le$ ADDITIVE GRAPHICAL NASH $\le$ $1/\exp(n)$-$2$NASH
\item $1/\exp(n)$-$k$NASH $\le$ END OF THE LINE $\le$ 2D-BROUWER $\le$ $n$D-BROUWER $\le$ $1/\mbox{poly}(n)$-$2$NASH
\end{OneLiners}

The reductions in the first chain are by \cite{LT82,DGP09}, \cite{Pap94,DGP09}, \cite{CD06a,DGP09} and \cite{DGP09}, respectively. The reductions in the second chain are by \cite{LT82,DGP09}, \cite{CD06b}, \cite{CDT06}, \cite{CDT06}, respectively. For an overview of these celebrated results see \cite{Rou10}. In comparison, our reduction can be written as:

\begin{OneLiners}
\item $\epsilon_k$-$k$NASH $\le$ $\epsilon_m$-LINEAR-NASH $\le$ $\epsilon_2$-$2$NASH
\end{OneLiners}
where $\epsilon_k,\epsilon_m,\epsilon_2$ can either be all $1/\exp(n)$ or all $1/\mbox{poly}(n)$. Note there is gap between the second chain of reductions and our results - the second chain achieves a stronger reduction from $1/\exp(n)$-$k$NASH to $1/\mbox{poly}(n)$-$2$NASH. Achieving a direct version of this result by \cite{CDT06} is an interesting open problem. Note also that our reduction from $\epsilon_m$-LINEAR-NASH to $\epsilon_2$-$2$NASH is somewhat similar to the reduction from ADDITIVE GRAPHICAL NASH to $1/\exp(n)$-$2$NASH, however our reduction does not require the input game to be bipartite nor does it limit the number of interactions per player.

Another open question is the complexity of $\epsilon_k$-$k$NASH
and $\epsilon_2$-$2$NASH for \emph{constant} values of $\epsilon_k,\epsilon_2$.
As a quasi-polynomial algorithm is known \cite{Alt94,LMM03}, these problems
are not believed to be PPAD-complete. The current state-of-the-art is a polynomial-time algorithm for $\epsilon_2$-$2$NASH where $\epsilon_2\approx 0.667$ \cite{KS08}. For finding $\epsilon_2$-Nash equilibrium rather than $\epsilon_2$-well-supported Nash equilibrium, there is an algorithm where $\epsilon_2\approx 0.339$ \cite{TS07}
(see also \cite{DMP07}, \cite{BBM07}, \cite{TS09}). On the negative side, several
algorithmic techniques have been ruled out \cite{HK09}, \cite{DP09}.

\paragraph{Reductions to 2-player games and linearization}

The empirical success of the Lemke-Howson algorithm \cite{LH64} for finding Nash equilibrium in 2-player games has motivated research on extending it to a more general class of games. Daskalakis et al.$\ $show a general reduction from \emph{succinct} games to 2-player games, which can be applied to any game in which the expected payoffs can be calculated using only ${+,*,\max}$ \cite{DFP06}. Their reduction goes through the steps of the first chain of reductions above. Govindan and Wilson present a non-polynomial linearizing reduction, which reduces multiplayer games to polymatrix games while preserving approximate Nash equilibrium \cite{GW10}. Their reduction introduces a central coordinator player, who interacts bilaterally with every player while simulating the combined behavior of the other players. 

In addition, linearization is also related to Etessami and Yannakakis's formulation of PPAD as the class of fixed-point problems for piecewise-linear functions (computable by ${+,\mbox{scale},\max}$) \cite{EY07}.

\section{A Linear Multiplication Gadget}
\label{sec:linear_mult}

In this section we construct a linear multiplication gadget using standard gadgets as building blocks. 

\begin{theorem}[Linear Multiplication Gadget]
\label{thm:exists-linear-mult}
There exist constants $\epsilon_{0}<1,c,d$ and an increasing polynomial function $f$ such that the following holds. For every $\epsilon<\epsilon_{0}$, there exists a linear multiplication gadget $G_{*}=G_*(\epsilon)$ of size $O(m\cdot f(\frac{1}{\epsilon}))$, such that in an $\epsilon$-well-supported Nash equilibrium, the output of $G_{*}$ equals the product of its $m$ inputs up to an additive error of $\pm dm\epsilon^{c}$.
\end{theorem}

We develop two different constructions of $G_*$, with two different sets of parameters $\epsilon_{0},c,d,f$.

\begin{lemma}[Polynomial-Sized Construction]
\label{lem:construction1}
\lref[Theorem]{thm:exists-linear-mult} holds with the following parameters:\footnote{The choice of $d=19$ simplifies the proof, but it is not hard to
show that for the same construction $d$ can be replaced with a smaller
value.} $\epsilon_{0}=\frac{1}{4}$, $c=1$, $d=19$ and $f(x)=x^{2}$.
\end{lemma}

\begin{lemma}[Logarithmic-Sized Construction]
\label{lem:construction2}
\lref[Theorem]{thm:exists-linear-mult} holds with the following parameters: $\epsilon_{0}=\frac{1}{10^5}$, $c=\frac{1}{2}$, $d=3$ and $f(x)=\log x$.
\end{lemma}

The second construction gives a smaller gadget with size $O(m\log\frac{1}{\epsilon})$ instead of $O(\frac{m}{\epsilon ^2})$, but is more complicated than the first construction. The rest of this section describes the first construction and proves \lref[Lemma]{lem:construction1}. Details of the second construction and the proof of \lref[Lemma]{lem:construction2} appear in \lref[Section]{sec:linear-mult-2}.

\subsection{Linear Gadgets}
\label{sec:linear-gadgets}

Goldberg and Papadimitriou developed the framework of gadgets \cite{GP06}, carefully-engineered games that simulate arithmetic calculations and are useful in many PPAD-completeness results (see, e.g., \cite{DGP09}). The players of a gadget game are typically \emph{binary}, representing numerical values in the range $[0,1]$. 

\begin{definition}[Binary Player]
\label{def:binary-player}
A \emph{binary player} $P$ is a player that has exactly two pure strategies 0 and 1. We say $P$ \emph{represents} the numerical value $p\in[0,1]$ if her mixed strategy is $(1-p,p)$, i.e. she plays pure strategy 1 with probability $p$.
\end{definition}

Gadget games have three kinds of binary players - one or more input players,\footnote{Gadgets can also have non-binary input players, in which case the value the input players represent is considered to be the probability with which they play a certain predetermined pure strategy.} one output player, and one or more auxiliary players. The \emph{size} of a gadget is the number of its auxiliary and output players. The values represented by the input and output players are the \emph{input values} and \emph{output value} of the gadget. In every $\epsilon$-well-supported Nash equilibrium of the gadget game, the output value $p$ is equal to the result of an arithmetic operation on the input values (up to small error). This arithmetic relation between the input and output values is the \emph{guarantee} of the gadget. To achieve the guarantee, the output player is incentivized to play the appropriate value $p$, by choosing appropriate payoff values for both the output and auxiliary players. Our reductions require gadgets with \emph{linear} guarantees, which differ slightly from the \emph{graphical} and \emph{additive}-graphical gadgets used in previous works.

\begin{definition}[Linear Gadgets]
\label{def:linear-gadget}A \emph{linear gadget} is a polymatrix gadget game with payoffs in $[0,1]$. Linear gadgets simulate linear arithmetic operations, i.e. their guarantee is a linear relation between the input and output values.
\end{definition}

Several gadgets can be combined into a single game, much like arithmetic gates are combined into a circuit to carry out involved calculations. By setting the output player $P$ of gadget $G_1$ to be an input player of gadget $G_2$, the value $p$ represented by $P$ is shared among the gadgets. We represent a combination of gadgets by a series of calculations on the input and output values. For example, if $P'$ is the output player of $G_2$ representing the value $p'$, then we write the above combination as $p'=G_2(p)$ (where in turn $p=G_1(\dots)$). The following fact explains why the same player can be an input player of multiple gadgets, but can only be the output player of a single gadget (as for auxliary players, they are considered part of the inner implementation and are thus never shared among different gadgets). It is a consequence of the gadget determining the payoff matrix of its output player, but not of its input players.

\begin{fact}[Combining Gadgets]
\label{fac:combining-gadgets}
For every game in which no player is the output player of more than one gadget, the guarantees of all gadgets hold simultanuously when the game is in $\epsilon$-well-supported Nash equilibrium.
\end{fact}

\subsubsection{Standard Gadgets}

The following gadgets are constructed by Daskalakis et al.$\ $\cite{DGP09}. To demonstrate the principle behind their construction, we include here the proof of \lref[Lemma]{lem:threshold-gadget}; proofs of \lref[Lemma]{lem:and-gadget} and \lref[Lemma]{lem:sum-gadget} appear in \lref[Appendix]{sec:additional-gadgets} for completeness.

\begin{lemma}[Linear Threshold Gadget]
\label{lem:threshold-gadget}
For every rational $\zeta\in\left[0,1\right]$, there exists a linear gadget $G_{>\zeta}$ of size $O(1)$ with input $p_1$, such that in an $\epsilon$-well-supported Nash equilibrium the output is 1 if $p_1>\zeta+\epsilon$ and 0 if $p_1<\zeta-\epsilon$, and otherwise it may be any value in $[0,1]$.
\end{lemma}

\begin{proof}
Let $P_1,P$ be the input and output players of $G_{>\zeta}$ representing values $p_1,p$, respectively. We set the payoff matrix $M^{P,P_1}$ to be:
$$
M^{P,P_1}=\left(\begin{array}{cc}
\zeta & \zeta\\
0 & 1\end{array}\right)
$$
$G_{>\zeta}$ has no auxiliary players, and so this concludes the construction. We now show that when $G_{>\zeta}$ is in $\epsilon$-well-supported Nash equilibrium, the guarantee of this gadget holds. Let $\vect{p^1}=(1-p_1,p_1)$ be player $P_1$'s mixed strategy in the $\epsilon$-well-supported Nash equilibrium. The expected payoff vector $\vect{u^P}$ of player $P$ is equal to:
$$
\vect{u^P}=M^{P,P_1}\vect{p^1}=(\zeta,p_1)
$$
If $p_1>\zeta+\epsilon$, the only $\epsilon$-best response for player $P$ is pure strategy 1, so $P$'s mixed strategy $(1-p,p)$ in the $\epsilon$-well-supported Nash equilibrium must be $(0,1)$ and thus $p=1$. Similarly, if $p_1<\zeta-\epsilon$ then $(1-p,p)=(1,0)$ and thus $p=0$.
\end{proof}

\begin{lemma}[Linear AND Gadget]
\label{lem:and-gadget}
There exists a linear gadget $G_{\wedge}$ of size $O(1)$ with inputs $p_1,p_2$, such that in an $\epsilon$-well-supported Nash equilibrium where $\epsilon<\frac{1}{4}$ the output is 1 if $p_1=p_2=1$ and 0 if $(p_1=0)\vee(p_2=0)$, and otherwise it may be any value in $[0,1]$.
\end{lemma}

\begin{lemma}[Linear Scaled-Summation Gadget]
\label{lem:sum-gadget}
For every rational $\zeta\in\left[0,1\right]$, there exists a linear gadget $G_{+,*\zeta}$ of size $O(1)$ with inputs $p_1,\dots,p_m$, such that in an $\epsilon$-well-supported Nash equilibrium the output is $\min\{\zeta(p_1+\dots+p_m),1\}\pm\epsilon$.
\end{lemma}

In addition, there exist standard gadgets for multiplication, but these are inherently nonlinear - the constructions are based on expected payoffs being multiplicative functions in players' probabilities (see, e.g., \cite{DGP09}).  

\subsection{Construction}

Here we show the construction of $G_*$ that will be used to prove \lref[Lemma]{lem:construction1}. We show a construction for multiplying 2 inputs, and multiplying $m$ inputs can be achieved by connecting $m-1$ copies of this construction serially. Let $P_1,P_2$ be the input players representing values $p_1,p_2$, and let $P$ be the output player representing value $p$. Let $\tau=3\epsilon$ (for simplicity assume that $1/\tau$ is integer). We first encode every input in unary representation, with precision of up to $\pm\tau$. For this we use $2/\tau$ auxiliary players: The vectors $\vect{v^1}=(v_1^1,\dots,v_{1/\tau}^1)$ and $\vect{v^2}=(v_1^2,\dots,v_{1/\tau}^2)$ of values represented by auxiliary players $\{V_i^1\}$ and $\{V_i^2\}$ are the unary encodings. The $i$'th unary bit of $p_1$ is $v_i^1$, and it is calculated by the threshold gadget $G_{>\zeta}$ (\lref[Lemma]{lem:threshold-gadget}) as follows: $v_i^1=G_{>i\tau}(p_1)$. Similarly, $v_i^2=G_{>i\tau}(p_2)$. Then we perform unary multiplication using the AND gadget $G_{\wedge}$ (\lref[Lemma]{lem:and-gadget}). The result is a matrix $U$, which contains $1/{\tau}^2$ values $u_{i,j}=G_\wedge(v_i^1,v_j^2)$, represented by auxiliary players $\{U_{i,j}\}$. The construction is complete by summing up and scaling $U$'s entries using the scaled-summation gadget $G_{+,*\zeta}$ (\lref[Lemma]{lem:sum-gadget}) as follows: $p=G_{+,*\tau^2}(u_{1,1},u_{1,2},\dots,u_{1/\tau,1/\tau})$. This establishes the relation between the input values $p_1,p_2$ and the output value $p$ of $G_*$. Note that the payoffs of all players are determined by the standard gadgets.

\subsection{Correctness}

We prove \lref[Lemma]{lem:construction1} for the case $m=2$. Namely, we show that for every $\epsilon<1/4$, when $G_*$ is in $\epsilon$-well-supported Nash equilibrium then $p=p_{1}p_2\pm d\epsilon$, that $G_*$ is linear and that the size of $G_*$ is $O(1/\epsilon^2)$. The proof of \lref[Lemma]{lem:construction1} for general $m$ follows, since concatenating $m-1$ copies of $G_*$ increases the error and gadget size by a multiplicative factor of $m$.

\begin{proofof}{\lref[Lemma]{lem:construction1} (Polynomial-Sized Construction)} First note that $G_*$ is a combination of linear gadgets and is thus itself linear. The size of $G_*$ is $O(1/{\tau}^2)$, the total size of the standard gadgets ($2/\tau$ threshold gadgets $G_{>\zeta}$, $1/{\tau}^2$ AND gadgets $G_{\wedge}$, and 1 scaled-summation gadget $G_{+,*\zeta}$, all of size $O(1)$). 

We assume $G_*$ is in $\epsilon$-well-supported Nash equilibrium where $\epsilon<1/4$, and write the input values $p_1,p_2$ as integer multiples of $\tau$ plus a small error: let $p_1=i^{*}\tau+\delta_1$ and $p_2=j^{*}\tau+\delta_2$, where $0\le i^*,j^*\le 1/\tau$ and $0\le\delta_1,\delta_2<\tau$. The following claim shows that the coefficients $i^*,j^*$ are correctly encoded as unary vectors $\vect{v^1},\vect{v^2}$, and is a direct consequence of the threshold gadget's guarantee (\lref[Lemma]{lem:threshold-gadget}). The threshold gadget is "brittle" in the sense that for a small range of inputs it returns an arbitrary output, but the choice of $\tau=3\epsilon$ ensures this happens for at most one unary bit.

\begin{myclaim}[Unary Encoding]
\label{cla:approx-i*}
$\vect{v^1}$ is of the form $(1,\dots,1,?,0,\dots,0)$, where $\norm{\vect{v^1}}=i^*\pm 1$ and '$?$' denotes any value in $[0,1]$. The same holds for $\vect{v^2}$ and $j^*$. 
\end{myclaim}

\begin{proof}
Consider the $i$'th entry of $\vect{v^1}$. By construction, $v_i^1=G_{>i\tau}(p_1)$. By \lref[Lemma]{lem:threshold-gadget}, $v_i^1$ indicates whether $p_1>i\tau+\epsilon$ or $p_1<i\tau-\epsilon$, and otherwise can be any value in $[0,1]$. Since $\tau=3\epsilon$ we know that $p_1>(i^*-1)\tau+\epsilon$, therefore for every $i\le i^*-1$ entry $v^1_i$ is equal to 1, and in total $\norm{\vect{v^1}}\ge i^*-1$. On the other hand we know that $p_1<(i^*+2)-\epsilon$, therefore for every $i\ge i^*+2$ entry $v^1_i$ is equal to 0, and in total $\norm{\vect{v^1}}\le i^*+1$. Moreover, since $\tau>2\epsilon$, there can be at most one value of $i$ for which $p_1=i\tau\pm\epsilon$, and so there can be at most one entry $i$ of $\vect{v^1}$ which is an arbitrary value '?' in $[0,1]$.
\end{proof}

The rest of the proof of \lref[Lemma]{lem:construction1} is a straightforward corollary of the other gadget guarantees. Let $\norm{U}$ denote the sum $\sum_{i,j}u_{i,j}$ of matrix $U$'s entries. By construction, $p=G_{+,*\tau^2}(u_{1,1},u_{1,2},\dots,u_{1/\tau,1/\tau})$, thus by the guarantee of gadget $G_{+,*\zeta}$ (\lref[Lemma]{lem:sum-gadget}), $p={\tau}^2\norm{U}\pm\epsilon$. We write the product $p_{1}p_2$ as an integer multiple of ${\tau}^2$ up to a small error:
$i^{*}j^{*}\tau^2\le p_{1}p_2<i^{*}j^{*}\tau^2+3\tau$. The next claim shows that $\norm{U}$ gives approximately the correct coefficient of ${\tau}^2$.

\begin{myclaim}
\label{cla:approx-i*j*}
$(i^*-1)(j^*-1)\le\norm{U}\le(i^*+1)(j^*+1)$
\end{myclaim}

\begin{proof}
Consider the $(i,j)$'th entry of $U$. By construction, $u_{i,j}=G_\wedge(v_i^1,v_j^2)$. By \lref[Lemma]{lem:and-gadget}, if $\epsilon<1/4$ and $v_i^1=v_j^2=1$ then $u_{i,j}=1$. By \lref[Claim]{cla:approx-i*}, there are at least $(i^*-1)(j^*-1)$ pairs $i,j$ such that $v_i^1=v_j^2=1$, and so $\norm{U}\ge(i^*-1)(j^*-1)$. Similarly, by \lref[Lemma]{lem:and-gadget}, if $\epsilon<1/4$ and $v_i^1=0\vee v_j^2=0$ then $u_{i,j}=0$. By \lref[Claim]{cla:approx-i*}, there are at most $(i^*+1)(j^*+1)$ pairs $i,j$ such that $v_i^1\ne 0\wedge v_j^2\ne 0$, and so $\norm{U}\le(i^*+1)(j^*+1)$.
\qed
\end{proof}

Since $i^*,j^*\le 1/\tau$, it follows from the above claim that $\norm{U}=i^*j^*\pm(2/\tau+1)$. So $p={\tau}^2i^*j^*\pm(3\tau+\epsilon)=p_1p_2\pm(6\tau+\epsilon)=p_1p_2\pm d\epsilon$, where $d=19$. This concludes the proof of \lref[Lemma]{lem:construction1}, showing that $G_*$  outputs the product of its inputs $p_{1}p_2$ up to a small error of $\pm d\epsilon$.
\end{proofof}

\begin{example}
\label{exa:mult-example}
Let $p_1=7\tau+\epsilon/4$ and $p_2=2\tau+(\tau-\epsilon/8)$. First we find the unary encoding: $\vect{v^1}=(1,1,1,1,1,1,?,0,\dots,0)$ and $\vect{v^2}=(1,1,?,0,\dots,0)$. Then we perform unary multiplication:
$$
U'=
\left(\begin{array}{cccccccc}
1 & 1 & 1 & 1 & 1 & 1 & ? & 0\\
1 & 1 & 1 & 1 & 1 & 1 & ? & 0\\
? & ? & ? & ? & ? & ? & ? & 0\\
0 & 0 & 0 & 0 & 0 & 0 & 0 & 0\\
\end{array}\right),
U=
\left(\begin{array}{cc}
U' & 0\\
0 & 0\\
\end{array}\right)_{1/\tau\times 1/\tau}
$$
Summing up and scaling the entries of $U$ we get $p=12{\tau}^2+O(\epsilon)$, which is close to $p_1p_2$ up to $O(\epsilon)$.
\end{example}

\section{Linearizing Multiplayer Games}
\label{sec:first-part-of-reduction}

In this section we show a direct reduction from $k$-player games to polymatrix games. Let $G_k$ denote the input game to the reduction, and let $G_m$ denote the corresponding output game. The reduction relies on the fact that, although $G_k$'s expected payoffs are nonlinear in its players' probabilities, they are linear in products of its players' probabilities. A key component of our reduction is a linear multiplication gadget for computing these products, which exists according to \lref[Theorem]{thm:exists-linear-mult}. Let $f$ be an increasing polynomial function as in \lref[Theorem]{thm:exists-linear-mult}. 

\begin{theorem}[A Linearizing Reduction]
\label{thm:first-reduction}
For every $\epsilon_k<1$, there exists a direct reduction from $\epsilon_k$-$k$NASH to $\epsilon_m$-LINEAR-NASH, where $\epsilon_m=\mbox{poly}(\epsilon_k/\abs{G_k})$. The reduction runs in polynomial time in $\abs{G_k}$ and in $f(1/\epsilon_k)$.
\end{theorem}

\begin{lemma}[Recovering $\epsilon_{k}$-Well-Supported Nash]
\label{lem:first-k-strategies}
Let $(\vect{p^1},\dots,\vect{p^m})$ be an $\epsilon_m$-well-supported
Nash equilibrium of $G_m$. Then the first $k$ mixed strategies $\vect{p^1},\dots,\vect{p^k}$ form an $\epsilon_k$-well-supported Nash equilibrium of $G_k$.
\end{lemma}

\subsection{Preserving Expected Payoffs}

The following lemma will be useful in desiging the linearizing reduction. Let $G_k$ be a game with $k$ players, $n$ pure strategies each. Let $G_m$ be a game with $m>k$ players, where the first $k$ players have the same pure strategies as the players of $G_k$. Let $(\vect{p^1},\dots,\vect{p^k}),(\vect{p^1},\dots,\vect{p^m})$ be mixed strategy profiles of $G_k,G_m$.

\begin{lemma}[Almost Equal Expected Payoffs]
\label{lem:equal-expected-payoffs}
If for every player $1\le i\le k$, the expected payoff vectors $\vect{u^i_{G_m}}$ and $\vect{u^i_{G_k}}$ are entry-wise equal up to an additive factor of $\delta$, and $(\vect{p^1},\dots,\vect{p^m})$ is an $\epsilon_m$-well-supported Nash equilibrium of $G_m$, then $(\vect{p^1},\dots,\vect{p^k})$ is an $\epsilon_k$-well-supported Nash equilibrium of $G_k$ where $\epsilon_k=2\delta+\epsilon_m$.
\end{lemma}

\begin{proof}
Let $j\in[n]$ be a pure strategy in the support of player $i$ ($p^i_j>0$). We know that $j$ is an $\epsilon_m$-best response in $G_m$. Assume for contradiction that $j$ is \emph{not} an $\epsilon_k$-best response in $G_k$, i.e. there is a pure strategy $j'\in[n],j'\ne j$ such that $\vect{u^i_{G_k}}(j')>\vect{u^i_{G_k}}(j)+\epsilon_k$. So 
$\vect{u^i_{G_m}}(j')+\delta>\vect{u^i_{G_m}}(j)-\delta+\epsilon_k$. Since $\epsilon_k-2\delta=\epsilon_m$, then $\vect{u^i_{G_m}}(j')>\vect{u^i_{G_m}}(j)+\epsilon_m$, contradiction. 
\qed
\end{proof}

\subsection{The Linearizing Reduction}

Given an input pair $(G_k,\epsilon_k)$, we find an output pair $(G_m,\epsilon_m)$ as follows. Let $\epsilon_0<1,c,d$ be the constant parameters of \lref[Theorem]{thm:exists-linear-mult}. Then $\epsilon_m=\min\{(\epsilon_k/3n^{k-1}dk)^{1/c},$ $\epsilon_0\}$. 
The players of $G_m$ are:
\begin{OneLiners}
\item \emph{Original players} - the first $k$ players of $G_m$ have the same pure strategies as $G_k$'s players. $\vect{p^i}$ denotes the mixed strategy of original player $i$. 
\item \emph{Mediator players} - for every $i\in[k]$, there is a set of $n^{k-1}$ binary players that corresponds to the set of $n^{k-1}$ pure strategy profiles of all original players except $i$. We denote by $Q_{\vect{s^{-i}}}$ the mediator player corresponding to pure strategy profile $\vect{s^{-i}}$ and by $q_{\vect{s^{-i}}}$ the represented value.
\item \emph{Gadget players} - all auxiliary players belonging to $kn^{k-1}$ copies of the linear multiplication gadget $G_*$.
\end{OneLiners}
Every mediator player is set to be the output player of a gadget $G_*$ as follows: $q_{\vect{s^{-i}}}=G_*(p^1_{\vect{s^{-i}}[1]},\dots,p^k_{\vect{s^{-i}}[k]})$. Thus, $q_{\vect{s^{-i}}}$ will be approximately equal to the probability with which the original players play the pure strategy profile $\vect{s^{-i}}$. Let $\vect{q^i}$ be the vector of values $\{q_{\vect{s^{-i}}}\}$, then it's approximately equal to $\vect{{\tilde{p}}^{-i}}$, the joint mixed strategy distribution of all original players except $i$. 

To complete the description of $G_m$ it remains to specify the non-zero payoff matrices of the original players (all other payoffs are determined by the gadgets). In $G_k$, the expected payoff vector of player $i$ is $u^i_{G_k}=M^i_{G_k}\vect{{\tilde{p}}^{-i}}$. In $G_m$, the payoff of original player $i$ will be influenced only by the $i$'th set of mediator players $\{Q_{\vect{s^{-i}}}\}$ who play $\vect{q^i}$. Instead of describing every payoff matrix $M^{i,Q_{\vect{s^{-i}}}}$ separately, (such a description appears in the proof of \lref[Lemma]{lem:first-k-strategies}), we describe one large payoff matrix $M^i_{G_m}$ that contains all the others (or more precisely, all their nonzero columns) as submatrices. We want the expected payoffs in $G_m$ to be as close as possible to those of $G_k$. Thus, we set $M^i_{G_m}=M^i_{G_k}$. This concludes the contruction.

\begin{figure}
\centering\includegraphics[scale=0.6]{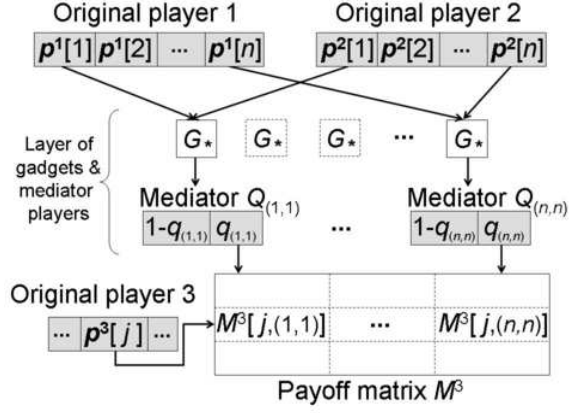}
\caption{Linearization of a 3-Player Game - Partial View of $G_m$} 
The arrows indicate how the probabilities of original players 1 and 2 influence the expected payoff of original player 3 via a layer of gadgets and mediator players.
\end{figure}

\subsection{Correctness}

\begin{proofof}{\lref[Theorem]{thm:first-reduction} (A Linearizing Reduction)}
The reduction runs in time polynomial in $\abs{G_k}=\Theta(kn^k)$ and in $f(1/\epsilon_k)$: The running time depends on the size of the polymatrix game $G_m$, which is polynomial in the number of its players. There are $k$ original players, $kn^{k-1}$ mediator players and $kn^{k-1}O(\abs{G_*})$ auxiliary players. By \lref[Theorem]{thm:exists-linear-mult}, $\abs{G_*}=O(k\cdot f(1/\epsilon_m))$. Since $f$ is a polynomial function and $\epsilon_m=\mbox{poly}(\epsilon_k/\abs{G_k})$, the total number of players is indeed polynomial in $\abs{G_k}$ and in $f(1/\epsilon_k)$. The rest of the proof follows from \lref[Lemma]{lem:first-k-strategies}.
\end{proofof}

\begin{proofof}{\lref[Lemma]{lem:first-k-strategies} (Recovering $\epsilon_k$-Well-supported Nash after Linearization)} Let $(\vect{p^1},\dots,\vect{p^m})$ be an $\epsilon_m$-well-supported
Nash equilibrium of $G_m$. We show that the first $k$ mixed strategies of $(\vect{p^1},\dots,\vect{p^m})$ form an $\epsilon_k$-well-supported Nash equilibrium of $G_k$. We would like to upper bound the entry-wise distance between the payoff vectors $u^i_{G_m},u^i_{G_k}$ so that we can apply \lref[Lemma]{lem:equal-expected-payoffs}. The proof proceeds as follows: We show that the expected payoff vector of original player $i$ in $G_m$ is $u^i_{G_m}=M^i_{G_m}\vect{q^i}$. Then we observe that the linear multiplication gadget $G_*$ guarantees that vectors $\vect{q^i}$ and $\vect{{\tilde{p}}^{-i}}$ are close to each other, and recall that $M^i_{G_m}=M^i_{G_k}$. Since all payoffs are in $[0,1]$, we conclude that the expected payoffs $u^i_{G_k}=M^i_{G_k}\vect{{\tilde{p}}^{-i}}$ are preserved in $G_m$. The proof of \lref[Lemma]{lem:first-k-strategies} is then immediate by preservation of expected payoffs (\lref[Lemma]{lem:equal-expected-payoffs}). 

We start by an alternative, more formal description of the original players' payoff matrices in $G_m$. Consider the payoff matrix $M^{i,Q_{\vect{s^{-i}}}}$, corresponding to the interaction between original player $i$ and mediator $Q_{\vect{s^{-i}}}$. Since $Q_{\vect{s^{-i}}}$ is a binary player with pure strategies $\{0,1\}$, the size of $M^{i,Q_{\vect{s^{-i}}}}$ is $n\times 2$. For every $j\in[n]$ we set $M^{i,Q_{\vect{s^{-i}}}}[j,1]=M^i_{G_k}[j,\vect{s^{-i}}]$ (where $M^i_{G_k}$ is the payoff matrix of player $i$ in game $G_k$), and $M^{i,Q_{\vect{s^{-i}}}}[j,0]=0$. So the column $M^{i,Q_{\vect{s^{-i}}}}[\cdot,0]$ corresponding to the mediator's pure strategy 0 is all-zeros. The payoff matrix $M^i_{G_m}$ was defined above to contain all nonzero columns of payoff matrices $M^{i,Q_{\vect{s^{-i}}}}$, i.e., all columns $M^{i,Q_{\vect{s^{-i}}}}[\cdot,1]$. It is now not hard to verify that $M^i_{G_m}=M^i_{G_k}$, and so the alternative description is equivalent to the original one.

\begin{myclaim}[Expected Payoffs Vector]
\label{cla:payoff-polymatrix}
For every $i\in[k]$, the expected payoff vector of original player $i$ in game $G_m$ is $u^i_{G_m}=M^i_{G_m}\vect{q^i}$.
\end{myclaim}

\begin{proof}
$u^i_{G_m}$ is equal to the sum of expected payoff vectors of original player $i$ from playing bilaterally against every mediator player in $\{Q_{\vect{s^{-i}}}\}$. Each expected payoff vector is a product of the payoff matrix $M^{i,Q_{\vect{s^{-i}}}}$ with vector $\vect{p^{Q_{\vect{s^{-i}}}}}=(1-q_{\vect{s^{-i}}},q_{\vect{s^{-i}}})$ (the mixed strategy played by the binary mediator player 
$Q_{\vect{s^{-i}}}$). By construction of $M^{i,Q_{\vect{s^{-i}}}}$, the expected payoff vector is equal to the product of column vector $M^{i,Q_{\vect{s^{-i}}}}[\cdot,1]$ with scalar $q_{\vect{s^{-i}}}$. Therefore, the sum of expected payoff vectors over all mediators is equal to $M^i_{G_m}\vect{q^i}$.
\qed
\end{proof}

We now show that $\vect{q^i}$ and $\vect{{\tilde{p}}^{-i}}$ are almost equal. Consider entry $q_{\vect{s^{-i}}}$ of $\vect{q^i}$. By construction, $q_{\vect{s^{-i}}}=G_*(p^1_{\vect{s^{-i}}[1]},\dots,p^k_{\vect{s^{-i}}[k]})$. 
By \lref[Theorem]{thm:exists-linear-mult} and since $\epsilon_m<\epsilon_0$, the gadget $G_*$ guarantees that $q_{\vect{s^{-i}}}=\prod_{i'\ne i}{p^{i'}_{\vect{s^{-i}}[i']}}\pm dk{(\epsilon_m)}^c$. By definition of $\vect{{\tilde{p}}^{-i}}$ as the joint mixed strategy distribution of all players except $i$ we get that $q_{\vect{s^{-i}}}=\vect{{\tilde{p}}^{-i}}[\vect{s^{-i}}]\pm dk{(\epsilon_m)}^c$. Thus, $\vect{q^i}=\vect{{\tilde{p}}^{-i}}\pm dk{(\epsilon_m)}^c$.

Using the fact that the entries of $M^i_{G_m},M^i_{G_k}$ are all in the range $[0,1]$, and that the dimensions of the matrices are $n\times n^{k-1}$, we conclude that $M^i_{G_m}\vect{q^i}=M^i_{G_k}\vect{{\tilde{p}}^{-i}}\pm n^{k-1}dk{(\epsilon_m)}^c$. We can now apply \lref[Lemma]{lem:equal-expected-payoffs} with $\delta=n^{k-1}dk{(\epsilon_m)}^c$. So $(\vect{p^1},\dots,\vect{p^k})$ is a $(2\delta+\epsilon_m)$-well-supported Nash equilibrium of $G_k$, and plugging in the chosen value of $\epsilon_m$ gives $\epsilon_k$-well-supported Nash equilibrium, as required. 
\end{proofof}

\section{Reducing Linear Games to Bimatrix Games}
\label{sec:second-part-of-reduction}

In this section we show how to replace the multiple players of a polymatrix game by two representative "super-players" of a bimatrix game. Let $G_m$ denote the input game to the reduction, and let $G_2$ denote the corresponding output game. 

\begin{theorem}[Linear to Bimatrix]
\label{thm:part2}
For every $\epsilon_m<1$, there exists a direct reduction from
$\epsilon_m$-LINEAR-NASH to $\epsilon_2$-$2$NASH, where $\epsilon_2=\mbox{poly}(\epsilon_m/\abs{G_m})$. The reduction runs in polynomial time in $\abs{G_m}$ and in $\log(1/\epsilon_m)$.
\end{theorem}

\begin{lemma}[Recovering $\epsilon_{m}$-Well-Supported Nash]
\label{lem:renormalize-recover}
For every $\epsilon_2$-well-supported Nash equilibrium $(\vect{x},\vect{y})$ of $G_2$, partitioning $\vect{y}$ into subvectors $\vect{y^1},\dots,\vect{y^m}$ of lengths $n_1,\dots,n_m$ and normalizing gives an $\epsilon_m$-well-supported Nash equilibrium $(\vect{y^1}/\norm{\vect{y^1}},$ $\dots,$ $\vect{y^m}/\norm{\vect{y^m}})$ of $G_m$.
\end{lemma}

\subsection{Imitation Games and Block $\epsilon$-Uniform Games}

The following definitions and lemmas will be useful in proving \lref[Theorem]{thm:part2}. An \emph{imitation game} is a bimatrix game in which both players have $N$ pure strategies, and the payoff matrix of player 2 is equal to the $N\times N$ identity matrix $I_N$. We call player 1 the \emph{leader} and player 2 the \emph{imitator}. A similar lemma to the following was proved in \cite{MT09} for the case of exact Nash equilibrium.

\begin{lemma}[Imitation]
\label{lem:imitation}
Let $(\vect{x},\vect{y})$ be an $\epsilon_2$-well-supported Nash equilibrium of an imitation game $G_2$ where $\epsilon_2\le 1/N$. Then $\mbox{support}(\vect{y})\subseteq\mbox{support}(\vect{x})$.
\end{lemma}

\begin{proof}
Assume pure strategy $j$ is not in $\mbox{support}(\vect{x})$, i.e. $x_j=0$. The expected payoff vector of the imitator is $\vect{u_{G_2}^{2}}=I_N\vect{x}=\vect{x}$, and so for pure strategy $j$ the expected payoff is $x_j=0$. Since $\vect{x}$ is a probability distribution vector with $N$ entries of which one is assumed to be zero, there exists a pure strategy $j'\ne j$ for which the imitator's expected payoff is $\vect{u_{G_2}^{2}}[j']=x_{j'}\ge 1/(N-1)>\epsilon$. The difference between the expected payoffs is more than $\epsilon$, so $j$ cannot be an $\epsilon$-best response for the imitator and so does not belong to $\mbox{support}(\vect{y})$. We conclude that $\mbox{support}(\vect{y})\subseteq\mbox{support}(\vect{x})$, as required.
\end{proof}

We call a bimatrix game \emph{block $\epsilon$-uniform} if player $1$'s payoff matrix $A$ is of the following form: 
\begin{OneLiners}
\item \emph{Block matrix:} $A$ is composed of $m^2$ blocks, where block $(i,i')$, denoted $A^{i,i'}$, is of size $n_i\times n_{i'}$; 
\item \emph{Very negative diagonal:} The $i$'th diagonal block $A^{i,i}$ is equal to $-\alpha E_{n_i}$, where $\alpha=8m^2/\epsilon$ and $E_{n_i}$ is the all-ones matrix of size $n_i\times n_i$; 
\item \emph{$[0,1]$ entries:} All other entries of $A$ are arbitrary values in the range $[0,1]$.
\end{OneLiners}

For a similar construction see the generalized Matching Pennies game of \cite{GP06}. If $\vect{x},\vect{y}$ is a mixed strategy profile of an $\epsilon$-block-uniform game, we denote by $\vect{x^1},\dots,\vect{x^m}$ and $\vect{y^1},\dots,\vect{y^m}$ the mixed strategy blocks of size $n_1,\dots,n_m$. We say that block $i$ belongs to the support of mixed strategy $\vect{x}$ if there is some pure strategy in block $i$ that belongs to this support. The following lemma shows that in a block $\epsilon$-uniform game, the weight of player 2 is $\epsilon$-uniformly divided among all blocks $i$ in $\mbox{support}(\vect{x})$.

\begin{lemma}[$\epsilon$-Uniform Weight Distribution]
\label{lem:uniform-weights}
Let $\vect{x},\vect{y}$ be an $\epsilon_2$-well-supported Nash equilibrium of a block $\epsilon_2$-uniform game $G_2$. If block $i\in[m]$ belongs to the support of $\vect{x}$, then for every $i'\in[m]$, $\norm{\vect{y^i}}\le\norm{\vect{y^{i'}}}+(1+\epsilon_2)/\alpha$.
\end{lemma}

\begin{proof}
The expected payoff vector $u^1_{G_2}$ of player 1 is $A\vect{y}$. By construction of matrix $A$, the expected payoff vector for playing pure strategies in block $i$ is $\sum_{i'\in[m]}{A^{i,i'}\vect{y^{i'}}}$. The domininant vector in this sum is $A^{i,i}\vect{y^i}$, whose entries are all $-\alpha\norm{\vect{y^i}}$. The entries of every other vector $A^{i,i'}\vect{y^{i'}}$ in the sum are in the range $[0,\norm{\vect{y^{i'}}}]$, and since $\vect{y}$ is a distribution vector, the total contribution to the sum is at most $\sum_{i'\in[m]}{\norm{\vect{y^{i'}}}}=1$. Thus, the expected payoff for playing any pure strategy in block $i$ is in the range $[-\alpha\norm{\vect{y^i}},-\alpha\norm{\vect{y^i}}+1]$. Assume for contradiction that $\norm{\vect{y^i}}>\norm{\vect{y^{i'}}}+(1+\epsilon_2)/\alpha$. Then the expected payoff for playing a pure strategy in block $i$ is at most $-\alpha(\norm{\vect{y^{i'}}}+(1+\epsilon_2)/\alpha)+1$, while the expected payoff for playing in block $i'$ is at least $-\alpha(\norm{\vect{y^{i'}}})$. The difference is more than $\epsilon_2$, contradicting the assumption that $i$ belongs to $\mbox{support}(\vect{x})$.
\qed
\end{proof}

If a game is both imitation and block $\epsilon$-uniform, then the weight of player 2 is divided $\epsilon$-uniformly among all blocks in $[m]$.

\begin{corollary}[Imitation and Block $\epsilon$-Uniform]
\label{cor:uniform-blocks}
Let $\vect{x},\vect{y}$ be an $\epsilon_2$-well-supported Nash equilibrium of a block $\epsilon_2$-uniform imitation game $G_2$, where $\epsilon_2\le 1/N$. Then for every two blocks $i,i'\in[m]$, $\norm{\vect{y^i}}=\norm{\vect{y^{i'}}}\pm (1+\epsilon_2)/\alpha$.
\end{corollary}

\begin{proof}
Since $\vect{y}$ is a distribution vector, there exists a block $i\in[m]$ such that $\norm{\vect{y^i}}\ge 1/m$. So $i$ belongs to the support of $\vect{y}$, and by \lref[Lemma]{lem:imitation}, $i$ also belongs to the support of $\vect{x}$. By \lref[Lemma]{lem:uniform-weights}, $1/m\le\norm{\vect{y^i}}\le\norm{\vect{y^{i'}}}+(1+\epsilon_2)/\alpha$ for every $i'\in[m]$. Since $(1+\epsilon_2)/\alpha<1/m$ we conclude that $0<\norm{\vect{y^{i'}}}$ for every $i'$. Thus by \lref[Lemma]{lem:imitation} all blocks are in $\mbox{support}(\vect{x})$ and get almost uniform weight.\qed
\end{proof}

\subsection{The Reduction}

Given an input pair $(G_m,\epsilon_m)$, we show how to find an output pair $(G_2,\epsilon_2)$, where $G_2$ has payoffs in the range $[-\alpha,1]$. To complete the reduction, $G_2$ can then be normalized by adding $\alpha$ to all payoffs and scaling by $1/(\alpha+1)$ ($\epsilon_2$ is also scaled). Let $N=\sum_{i=1}^m{n_i}$ be the total number of pure strategies in $G_m$. Let $\epsilon_2=\epsilon_m/N$. The pure strategies of every player in $G_2$ are the set $[N]$. The payoffs are chosen such that $G_2$ is both an imitation game and a block $\epsilon_2$-uniform game: 

$$
A=\left(\begin{array}{cccc}
-\alpha E_{n_1} & M^{1,2} & \cdots & M^{1,m}\\
 M^{2,1} & -\alpha E_{n_2} & & M^{2,m}\\
\vdots & & \ddots & \vdots\\
M^{m,1} & M^{m,2} & \cdots & -\alpha E_{n_m}
\end{array}\right)_{N\times N},
B=\left(\begin{array}{cccc}
I_{n_1} & 0 & \cdots & 0\\
0 & I_{n_2} & & 0\\
\vdots & & \ddots & \vdots\\
0 & 0 & \cdots & I_{n_m}
\end{array}\right)_{N\times N}
$$
where $M^{i,i'}$ is the payoff matrix of player $i$ for interacting with player $i'$ in $G_m$. 

\subsection{Correctness}
\label{sub:Correctness-part2}

\begin{proofof}{\lref[Theorem]{thm:part2} (Linear to Bimatrix)} First note that the reduction runs in time polynomial in $\abs{G_m}=\Theta(N^2)$ and in $\log(1/\epsilon_m)$: The running time depends on the size of the bimatrix game $G_2$, whose payoff matrices are of size $N^2$ with entries of size $O(\log\alpha)$. It's enough to prove \lref[Lemma]{lem:renormalize-recover} for the unnormalized game $G_2$ and $\epsilon_2=\epsilon_m/N$; this immediately gives a proof for $\epsilon_2=\epsilon_m/N(\alpha+1)$ after normalizing the payoffs from $[-\alpha,1]$ to $[0,1]$.\footnote{Note that every $\epsilon_m/N(\alpha+1)$-well-supported Nash equilibrium of the normalized game is an $\epsilon_m/N$-well-supported Nash equilibrium of the unnormalized game.} Since $\epsilon_m/N(\alpha+1)=\mbox{poly}(\epsilon_m/N)$, \lref[Theorem]{thm:part2} follows.

\end{proofof}

\begin{proofof}{\lref[Lemma]{lem:renormalize-recover} (Recovering $\epsilon_m$-Well-supported Nash from Bimatrix Game)} Let $(\vect{x},\vect{y})$ be an $\epsilon_2$-well-supported Nash equilibrium played in $G_2$, and let $(\vect{y^1}/\norm{\vect{y^1}},$ $\dots,$ $\vect{y^m}/\norm{\vect{y^m}})$ be a mixed strategy profile played in $G_m$. We show that this mixed strategy profile is actually an $\epsilon_m$-well-supported Nash equilibrium of $G_m$. 

For every player $i$ of $G_m$, we define an injective function $h_i:[n_i]\rightarrow[N]$ to be $h_i(j)=\sum_{i'<i}{n_{i'}}+j$. So $h_i$ maps the $j$'th pure strategy of player $i$ in $G_m$ to the $j$'th pure strategy in block $i$ of player 1 in $G_2$. We now show that player $i$'s expected payoff for playing $j$ in $G_m$ is closely related to player $1$'s expected payoff for playing $h_i(j)$ in $G_2$, assuming strategy profiles $(\vect{x},\vect{y})$ and $(\vect{y^1}/\norm{\vect{y^1}},$ $\dots,$ $\vect{y^m}/\norm{\vect{y^m}})$ are being played in $G_2$ and $G_m$, respectively. In fact, the expected payoffs are the same up to shifting by $\alpha\norm{\vect{y^i}}$ (the contribution from the diagonal of player 1's payoff matrix $A$), scaling by $m$ (the number of blocks on which $y$ is uniformly distributed), and small additive errors. As in \lref[Section]{sec:first-part-of-reduction}, the fact that the expected payoffs are preserved, even up to shift and scale, is enough for one game's $\epsilon$-well-supported Nash equilibrium to imply the other's.

\begin{myclaim}[Expected Payoffs are Preserved up to Shift and Scale]
\label{cla:relation-payoffs}
$\vect{u^i_{G_m}}[j] = m\cdot(\vect{u^1_{G_2}}[h_i(j)]+\alpha\norm{\vect{y^i}})\pm m^2(1+\epsilon_2)/\alpha$.
\end{myclaim}

\begin{proof}
By construction of matrix $A$, player 1's expected payoff vector for playing pure strategies in block $i$ is $A^{i,i}\vect{y^{i}}+\sum_{i'\ne i}{M^{i,i'}\vect{y^{i'}}}$. The entries of vector $A^{i,i}\vect{y^{i}}$ are $-\alpha\norm{\vect{y^i}}$, and the sum $\sum_{i'\ne i}{M^{i,i'}\vect{y^{i'}}}$ equals $\norm{\vect{y^{i'}}}\cdot\sum_{i'\ne i}{M^{i,i'}\vect{y^{i'}}}/\norm{\vect{y^{i'}}}= \norm{\vect{y^{i'}}}\cdot\vect{u^i_{G_m}}$. By \lref[Corollary]{cor:uniform-blocks}, and since $(\vect{x},\vect{y})$ in an $\epsilon_2$-well-supported Nash equilibrium of $G_2$, player 2's weight is distributed evenly over the $m$ blocks up to $(1+\epsilon_2)/\alpha$. It is not hard to see that \lref[Corollary]{cor:uniform-blocks} implies, for every $i'\in[m]$, that $\norm{\vect{y^{i'}}}=1/m\pm(1+\epsilon_2)/\alpha$. Plugging in, we get that entry $h_i(j)$ in player 1's expected payoff vector is $\vect{u^1_{G_2}}[h_i(j)]= -\alpha\norm{\vect{y^i}}+(1/m\pm(1+\epsilon_2)/\alpha)\cdot\vect{u^i_{G_m}}[j]$. The proof is complete by noting that player $i$'s expected payoff $\vect{u^i_{G_m}}[j]$ in $G_m$ is bounded by $m$, and by rearranging. 
\qed
\end{proof}

It's left to show that preservation of expected payoffs for playing $h_i(j)$ and $j$ up to shift and scale is enough to ensure that $(\vect{y^1}/\norm{\vect{y^1}},$ $\dots,$ $\vect{y^m}/\norm{\vect{y^m}})$ is an $\epsilon_m$-well-supported Nash equilibrium of $G_m$. More precisely, we show that if pure strategy $h_i(j)$ is an $\epsilon_2$-best response for player 1 in $G_2$, then pure strategy $j$ is an $\epsilon_m$-best response for player $i$ in $G_m$. We can then invoke \lref[Claim]{lem:imitation} by which player 2 only plays pure strategies that are $\epsilon_2$-best responses for player 1, and conclude that mixed strategy $\vect{y^i}/\norm{\vect{y^i}}$ contains only $\epsilon_m$-best responses for player $i$ in $G_m$.

Assume for contradiction that $j$ is not an $\epsilon_m$-best response for player $i$ in $G_m$. Then there exists another pure strategy $j'\in[m]$ such that $\vect{u^i_{G_m}}[j]<\vect{u^i_{G_m}}[j']-\epsilon_m$. But by \lref[Claim]{cla:relation-payoffs} this implies 
$\vect{u^1_{G_2}}[h_i(j)] < \vect{u^1_{G_2}}[h_i(j')]
+ 2m(1+\epsilon_2)/\alpha -\epsilon_m/m$. By choice of $\epsilon_2$ and $\alpha$, $\epsilon_2\le \epsilon_m/m -2m(1+\epsilon_2)/\alpha$. Thus, $h_i(j)$ cannot be an $\epsilon_2$-best response for player 1 in $G_2$, contradiction. This completes the proof of \lref[Lemma]{lem:renormalize-recover}.
\end{proofof}

\section{A Logarithmic-Sized Linear Multiplication Gadget}
\label{sec:linear-mult-2}

In this section we prove \lref[Lemma]{lem:construction2} by showing an alternative contruction of a linear multiplication gadget. The main difference from the construction shown in \lref[Section]{sec:linear_mult} is that the unary encoding is replaced by binary encoding. However, this introduces a new difficulty, since every gadget that performs binary bit extraction is inherently \emph{brittle}, i.e., its output is arbitrary for certain inputs. We use the bit extraction gadget of ~\cite{DGP09}, and overcome the brittleness using standard methods of averaging (somewhat simplified by introducing a new median gadget).

\subsection{Linear Gadgets}

We introduce several linear gadgets that will be useful for the construction. Additional gadgets that are known from previous works can be found in \lref[Appendix]{sec:additional-gadgets}. Throughout, we denote the input, output and auxiliary players of a gadget by $P_1,\dots,P_m,P,W,W_1,\dots,W_l$, and the values they represent by $p_1,\dots,p_m,p,w,w_1,\dots,w_l$.

The following gadget $G_{\text{mask}}$ treats its first input as a binary mask for its second input (i.e., performs multiplication between a binary input and an arbitrary input while maintaining linearity). Furthermore, it guarantees that if the second input is close to zero, the output will be close to zero as well. 

\begin{lemma}[Linear Mask Gadget]
\label{pro:correctness-of-Gmask}
There exists a linear gadget $G_{\text{mask}}$ such that in every $\epsilon$-well-supported Nash equilibrium:\[
p=\left\{ \begin{array}{ll}
p_{2}\pm\epsilon & \mbox{if }p_{1}=1\\
0 & \mbox{if }p_{1}=0\\
0\pm3\epsilon & \mbox{if }p_{2}=0\pm2\epsilon\end{array}\right.\]
\end{lemma}

\begin{proof}
Nonzero payoff matrices:
$$
M^{P,W}=\left(\begin{array}{cc}
1 & 0\\
0 & 1\end{array}\right),
M^{W,P}=\left(\begin{array}{cc}
0 & 1\\
0 & 0\end{array}\right),
M^{W,P_1}=\left(\begin{array}{cc}
2 & 0\\
0 & 0\end{array}\right),
M^{W,P_2}=\left(\begin{array}{cc}
0 & 0\\
0 & 1\end{array}\right)
$$
Expected payoff vectors: 
\begin{OneLiners}
\item $\vect{u^W}=M^{W,P}\vect{p^P}+M^{W,P_1}\vect{p^{P_1}}+M^{W,P_2}\vect{p^{P_2}} =(p+2(1-p_1),p_2)$;
\item $\vect{u^P}=M^{P,W}\vect{p^W}=(1-w,w)$.
\end{OneLiners}
Assume $G_{\text{mask}}$ is in $\epsilon$-well-supported Nash equilibrium. First we show that $p_1$ is a binary mask for $p_2$. If $p_1=0$, the only $\epsilon$-best response for player $W$ is 0, and so $w=0$ and also $p=0$ (actually this holds whenever $p_1<1/2$). If $p_1=1$, player $W$'s expected payoffs for playing strategies 0 and 1 are $p$ and $p_2$ respectively. We claim that $p$ must be equal to $p_2$ up to $\pm\epsilon$. Indeed, if $p>p_2+\epsilon$, $w=0$ and so $p=0$, contradiction. Similarly, if $p<p_2-\epsilon$, $w=1$ and so $p=1$, contradiction.

It is left to show that if $p_2$ is close to zero then $p$ is also close to zero. Assume for contradiction that $p_2\le 2\epsilon$ and $p>3\epsilon$. Pure strategy 1 must be an $\epsilon$-best response for $P$, therefore $w\ge(1-w)-\epsilon$. So pure strategy 1 must be an $\epsilon$-best response for $W$, therefore $p_2\ge p+2(1-p_1)-\epsilon$. Plugging in we get $2\epsilon\ge p_2\ge p+2(1-p_1)-\epsilon>3\epsilon+2(1-p_1)-\epsilon= 2\epsilon+2-2p_1$. This implies that $p_1>1$, contradiction. 
\end{proof}

\begin{lemma}[Linear Max Gadget]
\label{pro:correctness-of-Gmax}
There exists a linear gadget $G_{\max}$ such that in every $\epsilon$-well-supported Nash equilibrium, $p=\max\{p_1,p_2\}\pm 4\epsilon$.
\end{lemma}

\begin{proof}
The construction is by combining gadgets: \begin{eqnarray*}
w_1 & = & G_{<}\left(p_{1},p_{2}\right)\\
w_2 & = & G_{-}\left(p_{2},p_{1}\right)\\
w_3 & = & G_{\text{mask}}\left(w_{1},w_{2}\right)\\
p & = & G_{+,*1}\left(w_{3},p_{1}\right)\end{eqnarray*}
The correctness follows almost immediately from the guarantees of the combined gadgets (see \lref[Lemma]{lem:sum-gadget}, \lref[Lemma]{pro:correctness-of-Gmask}, \lref[Lemma]{pro:correctness-of-G<}, \lref[Lemma]{pro:correctness-of-Gminus}). The idea is to set the output $p$ to be approximately equal to $w_{1}\left(p_{2}-p_{1}\right)+p_{1}$, where $w_{1}$ is an indicator whether $p_{1}<p_{2}$. Assume $G_{\max}$ is in $\epsilon$-well-supported Nash equilibrium. If $p_{1}<p_{2}-\epsilon$ then $w_{1}=1$, and so $p$ is approximately equal to $p_{2}$. Similarly, if $p_{1}>p_{2}+\epsilon$ then $w_{1}=0$, and so $p$ is approximately equal to $p_{1}$. In the case that $\abs{p_{2}-p_{1}}\le\epsilon$, $w_{1}$ may receive any arbitrary value, but $w_{2}\le 2\epsilon$ and so by the guarantee of $G_{\text{mask}}$, $w_{3}\le 3\epsilon$. The product $w_{1}\left(p_{2}-p_{1}\right)$ is calculated by $G_{\text{mask}}$ in order to maintain the linearity of the construction. 
\end{proof}

\begin{lemma}[Linear Min Gadget]
\label{pro:correctness-of-Gmin}
There exists a linear gadget $G_{\min}$ such that in every $\epsilon$-well-supported Nash equilibrium, $p=\min\{p_1,p_2\}\pm 8\epsilon$.
\end{lemma}

\begin{proof}
The construction is by combining gadgets:\begin{eqnarray*}
w_{1} & = & G_{1-x}\left(p_{1}\right)\\
w_{2} & = & G_{1-x}\left(p_{2}\right)\\
w_{3} & = & G_{\max}\left(w_{1},w_{2}\right)\\
p & = & G_{1-x}\left(w_{3}\right)\end{eqnarray*}
The correctness follows immediately by the guarantees of the combined gadgets (see \lref[Lemma]{pro:correctness-of-Gmax}, \lref[Lemma]{pro:correctness-of-Gcomplementary}).
\end{proof}

\begin{lemma}[Linear Median Gadget]
\label{pro:correctness-of-Gmed}
There exists a linear gadget $G_{\text{median}}$ such that in every $\epsilon$-well-supported Nash equilibrium, $p=\text{median}\{p_1,p_2,p_3\}\pm 20\epsilon$.
\end{lemma}

\begin{proof}
The construction is by combining gadgets:\begin{eqnarray*}
w_{1} & = & G_{\max}\left(p_{1},p_{2}\right)\\
w_{2} & = & G_{\min}\left(p_{1},p_{2}\right)\\
w_{3} & = & G_{\min}\left(p_{3},w_{1}\right)\\
p & = & G_{\max}\left(w_{2},w_{3}\right)\end{eqnarray*}
Assume $G_{\text{median}}$ is in $\epsilon$-well-supported Nash equilibrium. We use the following notation to prove correctness: $s=\min\left\{ p_{1},p_{2},p_{3}\right\} $, $m=\text{median}\left\{ p_{1},p_{2},p_{3}\right\} $
and $l=\max\left\{ p_{1},p_{2},p_{3}\right\} $, such that $s\le m\le l$. The values $w_{2}$ and $w_{3}$ are equal to two of the three values $\left\{ p_{1},p_{2},p_{3}\right\} =\left\{ s,m,l\right\} $,
up to an error of $\pm12\epsilon$ introduced by the maximum and minimum
gadgets. To see this, notice that if $w_{1}=p_{i\in\left\{ 1,2\right\} }\pm4\epsilon$,
then $w_{2}=p_{j\in\left\{ 1,2\right\} ,j\neq i}\pm4\epsilon$ and
$w_{3}=\min\left\{ p_{3},p_{i}\pm4\epsilon\right\} \pm8\epsilon$
(by \lref[Lemma]{pro:correctness-of-Gmax} and \lref[Lemma]{pro:correctness-of-Gmin}).
Now, if $m<l-4\epsilon$, both $w_{2}$ and $w_{3}$ must be strictly smaller
than $l$, so the two values they are approximately equal to must
be $s$ and $m$. Taking their maximum therefore results in the median
value: $\max\left\{ w_{2},w_{3}\right\} = m\pm12\epsilon$, and so
$p = m\pm\left(12\epsilon+4\epsilon\right)$
(\lref[Lemma]{pro:correctness-of-Gmax}). If, however, $m\ge l-4\epsilon$,
then $w_{2}$ and $w_{3}$ are approximately equal to either
$\left\{ s,m\right\} $, $\left\{ s,l\right\} $ or $\left\{ m,l\right\} $.
Taking the maximum of $\left\{ w_{2},w_{3}\right\} $
therefore results in either $m\pm\left(12\epsilon+4\epsilon\right)$
or $l\pm\left(12\epsilon+4\epsilon\right)=m\pm20\epsilon$.
\end{proof}

\subsection{Brittle Construction}

We first construct a brittle multiplication gadget denoted by $\tilde{G}_{*}$. Let $\beta=\frac{1}{2}\log(1/\epsilon)$ be a parameter of the construction (assume for simplicity that $\beta$ is integral). The output $p$ of $\tilde{G}_{*}$ is guaranteed to be $p_{1}p_{2}\pm O(\sqrt{\epsilon})$, but only as long as the input $p_{1}$ is far enough from any integer multiple of $2^{-\beta}$. For every $1\le i\le\beta$, let $B_i,S_i,W_i$ be auxiliary players representing values $b_i,s_i,w_i$ respectively. $\tilde{G}_{*}$ sets $b_{1},\dots,b_{\beta}$ to be the $\beta$ most significant bits of input $p_1$ using the bit-extraction gadget $G_{\text{bit}}$ (\lref[Lemma]{pro:correctness-of-Gbit}). Then, $\tilde{G}_{*}$ calculates $p_{2}\sum_{i=1}^{\beta}\left(b_{i}2^{-i}\right)$, which equals $p_{2}\left(p_{1}\pm2^{-\beta}\right)$.
The calculations are carried out in the following order: 

\begin{OneLiners}
\item The values $\left\{ p_{2}2^{-i}\right\} _{i\in[\beta]}$ are
calculated using the scaling gadget $G_{*\zeta}$ (\lref[Lemma]{pro:correctness-of-Gmult-by-const}), i.e., $s_i=G_{*2^{-i}}(p_2)$.
\item For every $i\in[\beta]$, $p_22^{-i}$ is multiplied by the extracted
bit $b_i$ using the mask gadget $G_{\text{mask}}$ (\lref[Lemma]{pro:correctness-of-Gmask}), i.e., $w_i=G_{\text{mask}}(b_i,s_i)$.
\item The values $\{p_22^{-i}b_i\}_{i\in[\beta]}$ are summed up using the summation gadget $G_+=G_{+,*1}$ (\lref[Lemma]{lem:sum-gadget}), i.e., $p=G_+(w_1,\dots,w_\beta)$.
\end{OneLiners}

\subsection{Correctness of Brittle Construction}

\begin{lemma}
\label{pro:correctness-of-Gbrittle}
Let $p_1,p_2$ be the input values of $\tilde{G}_{*}$. If $p_{1}$ is $3\beta\epsilon$-far from every integer multiple of $2^{-\beta}$, then in every $\epsilon$-well-supported Nash equilibrium where $\epsilon\le\frac{1}{10^{3}}$, the output value $p$ equals $p_{1}p_{2}\pm 2\sqrt{\epsilon}$. The size of $\tilde{G}_{*}$ is $O(\beta)$.
\end{lemma}

\begin{proof}
First observe that since $\beta=\frac{1}{2}\log(1/\epsilon)$ (i.e.,
$2^{-\beta}=\sqrt{\epsilon}$) and $\epsilon\le\frac{1}{10^{3}}$, it
holds that $2\cdot 3\beta\epsilon<2^{-\beta}$, and so $p_{1}$ can indeed
be $3\beta\epsilon$-far from any integer multiple of $2^{-\beta}$. We write $p_1$ as $p_{1}=\sum_{i=1}^{\beta}b_{i}^{*}2^{-i}+\delta$, where $3\beta\epsilon<\delta<2^{-\beta}-3\beta\epsilon$. Assume $\tilde{G}_{*}$ is in $\epsilon$-well-supported Nash equilibrium where $\epsilon\le\frac{1}{10^{3}}$.
By \lref[Lemma]{pro:correctness-of-Gmask}, \lref[Lemma]{pro:correctness-of-Gbit} and \lref[Lemma]{pro:correctness-of-Gmult-by-const}, we know that for every $i\in[\beta]$: \begin{eqnarray*}
b_{i} & = & b_{i}^{*}\\
s_{i} & = & p_{2}2^{-i}\pm\epsilon\\
w_{i} & = & s_{i}b_{i}\pm\epsilon=p_{2}2^{-i} b_{i}^{*}\pm2\epsilon\end{eqnarray*}
By Lemma \ref{lem:sum-gadget} and the value of $\delta$: \begin{eqnarray*}
p & = & \min\left\{ 1,\sum_{i=1}^{\beta}w_{i}\right\} \pm\epsilon\\
 & = & \min\left\{ 1,\sum_{i=1}^{\beta}\left(p_{2}2^{-i} b_{i}^{*}\pm2\epsilon\right)\right\} \pm\epsilon\\
 & = & \min\left\{ 1,p_{2}\sum_{i=1}^{\beta}\left(2^{-i} b_{i}^{*}\right)\pm2\beta\epsilon\right\} \pm\epsilon\\
 & = & \min\left\{ 1,p_{2}\left(p_{1}-\delta\right)\pm2\beta\epsilon\right\} \pm\epsilon\\
 & = & p_{1} p_{2}\pm\left(2^{-\beta}+3\beta\epsilon\right)\end{eqnarray*}
Plugging in $\beta=\frac{1}{2}\log(1/\epsilon)$, we get that $p=p_{1}p_{2}\pm2\sqrt{\epsilon}$,
as required. Size of $\tilde{G}_{*}$: The bit-extraction gadget $G_{\text{bit}}$ requires $O(\beta)$ vertices (\lref[Lemma]{pro:correctness-of-Gbit}),
and the number of auxiliary vertices $\{S_{i}\}$
and $\left\{ W_{i}\right\}$ is also $O(\beta)$.
The other gadgets are of constant size. 
\end{proof}

\subsection{Robust Construction}

When $p_{1}$ is close to a multiple of $2^{-\beta}$, $\tilde{G}_{*}$'s output may be arbitrary. To circumvent this issue, the ultimate multiplication gadget $G_{*}$ applies $\tilde{G}_{*}$ three times, each time with a slightly perturbed copy of the input $p_{1}$. The perturbation guarantees that at most one of the three copies of $p_{1}$ is close to an integer multiple of $2^{-\beta}$, so that at least two of $\tilde{G}_{*}$'s three outputs are approximately correct. The difficulty is that we don't know which of the three outputs is approximately correct and which is arbitrary. We overcome this difficulty by taking the median of the three outputs as the final result, which is now guaranteed to be approximately equal to the required output $p_{1}p_{2}$.

The inputs to $\tilde{G}_{*}$ are set to be (up to $\pm O(\epsilon)$): $(\tilde{p_1},p_2)$, $(\tilde{p_1}-\Delta,p_2)$
and $(\tilde{p_1}-2\Delta,p_2)$, where $\tilde{p_1}=\max\{p_1,2\Delta\}$
and $\Delta=7\beta\epsilon$. This is achieved by combining the following gadgets:

\begin{OneLiners}
\item First, the value of $\tilde{p_1}$ is set: $c_1=G_{:=}(2\Delta+7\epsilon),
\tilde{p_1}=G_{\max}(p_1,c_1)$.
\item Then, two additional inputs are prepared: $c_2=G_{:=}(\Delta),c_3=G_{:=}(2\Delta), d_1=G_{-}(\tilde{p_1},c_2),d_2=G_{-}(\tilde{p_1},c_3)$.
\item $\tilde{G}_{*}$ is applied: $w_1=\tilde{G}_{*}(\tilde{p}_{1},p_{2}),
w_2=\tilde{G}_{*}(d_{1},p_{2}),
w_3=\tilde{G}_{*}(d_{2},p_{2})$.
\item The median is found: $p=G_{\text{median}}(w_1,w_2,w_3)$.
\end{OneLiners}

\subsection{Correctness of Robust Construction}

\begin{proofof}{\lref[Lemma]{lem:construction2} (Logarithmic-Sized Construction)} First note that $G_*$ is a combination of linear gadgets and is thus itself linear. The size of $G_{*}$ is $O(\beta)=O(\log\frac{1}{\epsilon})$, since the brittle multiplication gadget $\tilde{G}_{*}$ requires $O(\beta)$ vertices (\lref[Lemma]{pro:correctness-of-Gbrittle}), and the number of other auxiliary vertices is constant. 

Assume $G_{*}$ is in $\epsilon$-well-supported Nash equilibrium where $\epsilon\le\frac{1}{10^{5}}$.
By the gadget guarantees we know that
$\tilde{p}_{1}\ge 2\Delta+2\epsilon$, and that $d_{1}=\tilde{p}_{1}-\Delta\pm2\epsilon$ and $d_{2}=\tilde{p}_{1}-2\Delta\pm2\epsilon$ (\lref[Lemma]{pro:correctness-of-Gmax}, \lref[Lemma]{pro:correctness-of-Gminus} and \lref[Lemma]{pro:correctness-of-Gassign}).
Since $\epsilon<\frac{1}{10^{5}},\Delta=7\beta\epsilon$ and $\beta=\frac{1}{2}\log\frac{1}{\epsilon}$,
it can easily be verified that $\tilde{p}_{1}>d_{1}>d_{2}\ge0$ and
that the distance between each consecutive pair is $\Delta\pm4\epsilon$.

\begin{claim}
\label{cla:At-most-one}
At most one of $\tilde{p}_{1},d_{1},d_{2}$
can be $3\beta\epsilon$-close to a multiple of $2^{-\beta}$.
\end{claim}

\begin{proof}
Since the distance $\Delta\pm4\epsilon$ is larger than $2\cdot 3\beta\epsilon$,
if one of $\tilde{p}_{1},d_{1},d_{2}$ is $3\beta\epsilon$-close to a
certain multiple $k2^{-\beta}$, then the other two must be $3\beta\epsilon$-far
from $k2^{-\beta}$. Furthermore, since the distance is smaller than
$(2^{-\beta}-2\cdot 3\beta\epsilon)/2$, the other two must be $3\beta\epsilon$-far
from the nearby multiples $\left(k-1\right)2^{-\beta}$ and $\left(k+1\right)2^{-\beta}$
as well. 
\end{proof}

By \lref[Lemma]{pro:correctness-of-Gbrittle} and \lref[Claim]{cla:At-most-one}, at most one of $w_{1},w_{2},w_{3}$ can be arbitrary. There are two cases:

\begin{OneLiners}
\item The median is not the arbitrary value. Assume without loss of generality that the median is $w_{3}$ (since it is the furthest from $\tilde{p}_{1}p_{2}$).
By \lref[Lemma]{pro:correctness-of-Gmed} and \lref[Lemma]{pro:correctness-of-Gbrittle}:
\begin{eqnarray*}
p & = & w_{3}\pm20\epsilon\\
 & = & d_{2}p_{2}\pm\left(2\sqrt{\epsilon}+20\epsilon\right)\\
 & = & \tilde{p}_{1}p_{2}\pm\left(2\sqrt{\epsilon}+2\Delta+22\epsilon\right)\end{eqnarray*}

\item The median is the arbitrary value. Assume without loss of generality that the non-arbitrary
values are $w_{2}$ and $w_{3}$ (the furthest from $\tilde{p}_{1}p_{2}$).
The median is between these values, so we may assume without loss of generality that it
is equal to $w_{3}$, and proceed as in the previous case.
\end{OneLiners}
We have seen that in both cases, $p$ is close to $\tilde{p}_{1}p_{2}$.
It is now left to verify that $\tilde{p}_{1}$ is close to $p_{1}$.

\begin{claim}
$\tilde{p}_{1}=p_{1}\pm\left(2\Delta+11\epsilon\right)$.
\end{claim}

\begin{proof}
If $\max\left\{ p_{1},c_{1}\right\} =p_{1}$ then $\tilde{p}_{1}=p_{1}\pm4\epsilon$
(\lref[Lemma]{pro:correctness-of-Gmax}). Otherwise, $0\le p_{1}\le c_{1}=2\Delta+6\epsilon\pm\epsilon$
(\lref[Lemma]{pro:correctness-of-Gassign}) and $\tilde{p}_{1}=c_{1}\pm4\epsilon$. 
\end{proof}

We conclude that $p=p_{1}p_{2}\pm\left(2\sqrt{\epsilon}+4\Delta+37\epsilon\right)=p_{1}p_{2}\pm3\sqrt{\epsilon}$, as required.
\end{proofof}

\bibliography{bib}

\begin{thebibliography}{LMM03}

\bibitem[Alt94]{Alt94}
I.~Althofer.
\newblock On sparse approximations to randomized strategies and convex
  combinations.
\newblock {\em Linear Algebra and its Applications}, 240:9--19, 1994.

\bibitem[BBM07]{BBM07}
H.~Bosse, J.~Byrka, and E.~Markakis.
\newblock New algorithms for approximate {N}ash equilibria in bimatrix games.
\newblock In {\em WINE}, 2007.

\bibitem[Bub79]{Bub79}
V.~Bubelis.
\newblock On equilibria in finite games.
\newblock {\em International Journal of Game Theory}, 8(2):65--79, 1979.

\bibitem[CD06a]{CD06b}
X.~Chen and X.~Deng.
\newblock On the complexity of 2d discrete fixed point problem.
\newblock In {\em ICALP}, 2006.

\bibitem[CD06b]{CD06a}
X.~Chen and X.~Deng.
\newblock Settling the complexity of 2-player {N}ash-equilibrium.
\newblock In {\em FOCS}, 2006.

\bibitem[CDT06]{CDT06}
X.~Chen, X.~Deng, and S.~Teng.
\newblock Computing {N}ash equilibria: Approximation and smoothed complexity.
\newblock In {\em FOCS}, 2006.

\bibitem[DFP06]{DFP06}
C.~Daskalakis, A.~Fabrikant, and C.~H. Papadimitriou.
\newblock The game world is flat: The complexity of {N}ash equilibria in
  succinct games.
\newblock In {\em ICALP}, 2006.

\bibitem[DGP09]{DGP09}
C.~Daskalakis, P.~W. Goldberg, and C.~H. Papadimitriou.
\newblock The complexity of computing a {N}ash equilibrium.
\newblock {\em SIAM Journal on Computing}, 39(1):195--259, 2009.

\bibitem[DMP06]{DMP07}
C.~Daskalakis, A.~Mehta, and C.~H. Papadimitriou.
\newblock Progress in approximate {N}ash equilibria.
\newblock In {\em EC}, 2006.

\bibitem[DP09]{DP09}
C.~Daskalakis and C.~H. Papadimitriou.
\newblock On oblivious ptas's for {N}ash equilibrium.
\newblock In {\em STOC}, 2009.

\bibitem[EY07]{EY07}
K.~Etessami and M.~Yannakakis.
\newblock On the complexity of {N}ash equilibria and other fixed points.
\newblock In {\em FOCS}, 2007.

\bibitem[GKT50]{GKT50}
D.~Gale, H.~W. Kuhn, and A.~W. Tucker.
\newblock On symmetric games.
\newblock In H.~W. Kuhn and A.~W. Tucker, editors, {\em Contributions to the
  Theory of Games}, pages 81--87. Princeton, 1950.

\bibitem[GP06]{GP06}
P.~W. Goldberg and C.~H. Papadimitriou.
\newblock Reducibility among equilibrium problems.
\newblock In {\em STOC}, 2006.

\bibitem[GW10]{GW10}
S.~Govindan and R.~Wilson.
\newblock A decomposition algorithm for n-player games.
\newblock {\em Economic Theory}, 42(1):97--117, 2010.

\bibitem[HK09]{HK09}
E.~Hazan and R.~Krauthgamer.
\newblock How hard is it to approximate the best {N}ash equilibrium?
\newblock In {\em SODA}, 2009.

\bibitem[KS10]{KS08}
S.~C. Kontogiannis and P.~G. Spirakis.
\newblock Well supported approximate equilibria in bimatrix games.
\newblock {\em Algorithmica}, 57(4):653--667, 2010.

\bibitem[LH64]{LH64}
C.~E. Lemke and J.~T. Howson.
\newblock Equilibrium points of bimatrix games.
\newblock {\em SIAM Journal of Applied Mathematics}, 12:413--423, 1964.

\bibitem[LMM03]{LMM03}
R.~Lipton, E.~Markakis, and A.~Mehta.
\newblock Playing large games using simple strategies.
\newblock In {\em EC}, 2003.

\bibitem[MT09]{MT09}
A.~McLennan and R.~Tourky.
\newblock Imitation games and computation.
\newblock {\em Games and Economic Behavior}, 2009.

\bibitem[Nas51]{Nas51}
J.~F. Nash.
\newblock Non-cooperative games.
\newblock {\em Annals of Mathematics}, 54:289--295, 1951.

\bibitem[Pap94]{Pap94}
C.~H. Papadimitriou.
\newblock On the complexity of the parity argument and other inefficient proofs
  of existence.
\newblock {\em Journal of Computer and System Sciences}, 48(3):498--532, 1994.

\bibitem[Pap07]{Pap07}
C.~H. Papadimitriou.
\newblock The complexity of finding {N}ash equilibria.
\newblock In N.~Nisan, T.~Roughgarden, E.~Tardos, and V.~V. Vazirani, editors,
  {\em Algorithmic Game Theory}, chapter~2. Cambridge University Press, 2007.

\bibitem[Rou10]{Rou10}
T.~Roughgarden.
\newblock Computing equilibria: A computational complexity perspective.
\newblock {\em Economic Theory}, 42(1):193--236, 2010.

\bibitem[TS07]{TS07}
H.~Tsaknakis and P.~G. Spirakis.
\newblock An optimization approach for approximate {N}ash equilibria.
\newblock In {\em WINE}, 2007.

\bibitem[TS09]{TS09}
H.~Tsaknakis and P.~G. Spirakis.
\newblock A graph spectral approach for computing approximate {N}ash
  equilibria.
\newblock In {\em ECCC}, 2009.

\bibitem[vdLT82]{LT82}
G.~van~der Laan and A.~J.~J. Talman.
\newblock On the computation of fixed points in the product space of unit
  simplices and an application to noncooperative n-person games.
\newblock {\em Mathematics of Operations Research}, 7(1):1--13, 1982.

\bibitem[vS07]{vSte07}
B.~von Stengel.
\newblock Equilibrium computation for two-player games in strategic and
  extensive form.
\newblock In N.~Nisan, T.~Roughgarden, E.~Tardos, and V.~V. Vazirani, editors,
  {\em Algorithmic Game Theory}, chapter~3. Cambridge University Press, 2007.

\end{thebibliography}

\appendix
\section{Standard Gadgets}
\label{sec:additional-gadgets}

The following gadgets are constructed by Daskalakis et al.$\ $\cite{DGP09}. We denote the input and output players by $P_1,P_2,P$, and the values they represent by $p_1,p_2,p$. 

\begin{proofof}{\lref[Lemma]{lem:and-gadget} (Linear AND Gadget)}

Nonzero payoff matrices:
$$
M^{P,P_1}=M^{P,P_2}=\left(\begin{array}{cc}
\frac{3}{16} & \frac{3}{16}\\
0 & \frac{1}{2}\end{array}\right)
$$
Expected payoff vectors:
\begin{OneLiners}
\item $\vect{u^P}=M^{P,P_1}\vect{p^1}+M^{P,P_2}\vect{p^2}=(3/4,(p_1+p_2)/2)$.
\end{OneLiners}
Assume $G_{>\zeta}$ is in $\epsilon$-well-supported Nash equilibrium where $\epsilon<1/4$. If $p_1=p_2=1$, the only $\epsilon$-best response for player $P$ is pure strategy 1, so $p=1$. Similarly, if $(p_1=0)\vee(p_2=0)$ then $p=0$.
\end{proofof}

\begin{proofof}{\lref[Lemma]{lem:sum-gadget} (Linear Scaled-Summation Gadget)}

Let $P_1,\dots,P_m,P,W$ be the input players, output player and auxiliary player of $G_{+,*\zeta}$ respectively, representing values $p_1,\dots,p_m,p,w$. Nonzero payoff matrices:
$$
M^{W,P}=\left(\begin{array}{cc}
0 & 1\\
0 & 0\end{array}\right),
M^{W,P_i}=\left(\begin{array}{cc}
0 & 0\\
0 & \zeta\end{array}\right),
M^{P,W}=\left(\begin{array}{cc}
1 & 0\\
0 & 1\end{array}\right)
$$
Expected payoff vectors:
\begin{OneLiners}
\item $\vect{u^W}=M^{W,P}\vect{p^P}+\sum_{i\in[m]}{M^{W,P_i}\vect{p^{P_i}}}=(p,\zeta\sum_{i\in[m]}{p_i})$;
\item $\vect{u^P}=M^{P,W}\vect{p^W}=(1-w,w)$.
\end{OneLiners}
Assume $G_{+,*\zeta}$ is in $\epsilon$-well-supported Nash equilibrium. If player $W$ plays full support ($0<w<1$), then both of $W$'s pure strategies 0 and 1 must be $\epsilon$-best responses, and so $p=\vect{u^W}[0]=\vect{u^W}[1]\pm\epsilon=\zeta\sum_{i\in[m]}{p_i}\pm\epsilon$, as required. If $w=0$, the only $\epsilon$-best response for player $P$ is pure strategy 0, so $p=0$. Similarly, if $w=1$ then $p=1$. Case analysis:
\begin{OneLiners}
\item $\epsilon<\zeta\sum_{i\in[m]}{p_i}<1-\epsilon$: Assume for contradiction that $W$ does not play full support. Without loss of generality, assume $w=1$. But then $p=1$ and $\vect{u^W}[0]=1>\zeta\sum_{i\in[m]}{p_i}+\epsilon=\vect{u^W}[1]+\epsilon$, contradiction. Similarly, $w=0$ leads to contradiction. 
\item $\zeta\sum_{i\in[m]}{p_i}\le\epsilon$: Player $W$ can either play full support or pure strategy 0 (if $w=1$ then $p=1$ and the only $\epsilon$-best response for $W$ is pure strategy 0, contradiction). If $w=0$ then $p=0=\zeta\sum_{i\in[m]}{p_i}\pm\epsilon$, as required. 
\item $\zeta\sum_{i\in[m]}{p_i}\ge 1-\epsilon$: Similarly to the previous case, $W$ can either play full support or pure strategy 1, and if $w=1$ then $p=1=\min\{1,\zeta\sum_{i\in[m]}{p_i}\}\pm\epsilon$, as required. 
\end{OneLiners}
\end{proofof}

\begin{lemma}[Linear Comparison Gadget]
\label{pro:correctness-of-G<}
There exists a linear comparison gadget $G_{<}$ of size $O(1)$, such that in every $\epsilon$-well-supported Nash equilibrium, $p=1$ if $p_1<p_2-\epsilon$ and $p=0$ if $p_1>p_2+\epsilon$. 
\end{lemma}

\begin{proof}
Nonzero payoff matrices:
$$
M^{P,P_1}=\left(\begin{array}{cc}
0 & 1\\
0 & 0\end{array}\right),
M^{P,P_2}=\left(\begin{array}{cc}
0 & 0\\
0 & 1\end{array}\right)
$$
Expected payoff vectors:
\begin{OneLiners}
\item $\vect{u^P}=M^{P,P_1}\vect{p^1}+M^{P,P_2}\vect{p^2}=(p_1,p_2)$.
\end{OneLiners}
Assume $G_<$ is in $\epsilon$-well-supported Nash equilibrium. If $p_1<p_2-\epsilon$, the only $\epsilon$-best response for player $P$ is pure strategy 1, so $p=1$. Similarly, if $p_1>p_2+\epsilon$ then $p=0$.
\end{proof}

\begin{lemma}[Linear Minus Gadget]
\label{pro:correctness-of-Gminus}
There exists a linear subtraction gadget $G_{-}$ of size $O(1)$, such that in every $\epsilon$-well-supported Nash equilibrium, $p=\max\left\{ 0,p_{2}-p_{1}\right\} \pm\epsilon$.
\end{lemma}

\begin{proof}
Nonzero payoff matrices:
$$
M^{W,P}=\left(\begin{array}{cc}
0 & 1\\
0 & 0\end{array}\right),
M^{W,P_1}=\left(\begin{array}{cc}
0 & 0\\
0 & -1\end{array}\right),
M^{W,P_2}=\left(\begin{array}{cc}
0 & 0\\
0 & 1\end{array}\right),
M^{P,W}=\left(\begin{array}{cc}
1 & 0\\
0 & 1\end{array}\right)
$$
Expected payoff vectors:
\begin{OneLiners}
\item $\vect{u^W}=M^{W,P}\vect{p^P}+M^{W,P_1}\vect{p^1}+M^{W,P_2}\vect{p^2}=(p,p_2-p_1)$;
\item $\vect{u^P}=M^{P,W}\vect{p^W}=(1-w,w)$.
\end{OneLiners}
Assume $G_-$ is in $\epsilon$-well-supported Nash equilibrium. As in the proof of \lref[Lemma]{lem:sum-gadget}, it is not hard to show that either player $W$ plays full support (so both of $W$'s pure strategies must be $\epsilon$-best responses and $p=p_2-p_1\pm\epsilon$), or one of the following happens:
\begin{OneLiners}
\item $p_2-p_1>1-\epsilon$: Player $W$ can play pure strategy 1, and then $p=1=p_2-p_1\pm\epsilon$, as required. 
\item $p_2-p_1<\epsilon$: Player $W$ can play pure strategy 0, and then $p=0=\max\{0,p_2-p_1\}\pm\epsilon$, as required. 
\end{OneLiners}
\end{proof}

\begin{lemma}[Linear Complementary Gadget]
\label{pro:correctness-of-Gcomplementary}
There exists a linear complementary gadget $G_{1-x}$ of size $O(1)$, such that in every $\epsilon$-well-supported Nash equilibrium, $p=1-p_{1}\pm\epsilon$.
\end{lemma}

\begin{proof}
Nonzero payoff matrices:
$$
M^{W,P}=\left(\begin{array}{cc}
0 & 1\\
0 & 0\end{array}\right),
M^{W,P_1}=\left(\begin{array}{cc}
0 & 0\\
1 & 0\end{array}\right),
M^{P,W}=\left(\begin{array}{cc}
1 & 0\\
0 & 1\end{array}\right)
$$
Expected payoff vectors:
\begin{OneLiners}
\item $\vect{u^W}=M^{W,P}\vect{p^P}+M^{W,P_1}\vect{p^1}=(p,1-p_1)$;
\item $\vect{u^P}=M^{P,W}\vect{p^W}=(1-w,w)$.
\end{OneLiners}
Assume $G_{1-x}$ is in $\epsilon$-well-supported Nash equilibrium. As in the proof of \lref[Lemma]{lem:sum-gadget}, it is not hard to show that either player $W$ plays full support (so both of $W$'s pure strategies must be $\epsilon$-best responses and $p=1-p_1\pm\epsilon$), or one of the following happens:
\begin{OneLiners}
\item $1-p_1>1-\epsilon$: Player $W$ can play pure strategy 1, and then $p=1=1-p_1\pm\epsilon$, as required. 
\item $1-p_1<\epsilon$: Player $W$ can play pure strategy 0, and then $p=0=1-p_1\pm\epsilon$, as required. 
\end{OneLiners}
\end{proof}

\begin{lemma}[Linear Assignment Gadget]
\label{pro:correctness-of-Gassign}
For every rational $\zeta\in\left[0,1\right]$, there exists a linear assignment gadget $G_{:=\zeta}$ of size $O(1)$, such that in every $\epsilon$-well-supported Nash equilibrium, $p=\zeta\pm\epsilon$.
\end{lemma}

\begin{proof}
Nonzero payoff matrices:
$$
M^{W,P}=\left(\begin{array}{cc}
0 & 1\\
\zeta & \zeta\end{array}\right),
M^{P,W}=\left(\begin{array}{cc}
1 & 0\\
0 & 1\end{array}\right)
$$
Expected payoff vectors:
\begin{OneLiners}
\item $\vect{u^W}=M^{W,P}\vect{p^P}=(p,\zeta)$;
\item $\vect{u^P}=M^{P,W}\vect{p^W}=(1-w,w)$.
\end{OneLiners}
Proof of correctness as in \lref[Lemma]{pro:correctness-of-Gcomplementary}.
\end{proof}

\begin{lemma}[Linear Scaling Gadget]
\label{pro:correctness-of-Gmult-by-const}
For every rational $\zeta\in\left[0,1\right]$, there exists a linear scaling gadget $G_{*\zeta}$ of size $O(1)$, such that in every $\epsilon$-well-supported Nash equilibrium, $p=\zeta p_{1}\pm\epsilon$.
\end{lemma}

\begin{proof}
Nonzero payoff matrices:
$$
M^{W,P}=\left(\begin{array}{cc}
0 & 1\\
0 & 0\end{array}\right),
M^{W,P_1}=\left(\begin{array}{cc}
0 & 0\\
0 & \zeta\end{array}\right),
M^{P,W}=\left(\begin{array}{cc}
1 & 0\\
0 & 1\end{array}\right)
$$
Expected payoff vectors:
\begin{OneLiners}
\item $\vect{u^W}=M^{W,P}\vect{p^P}+M^{W,P_1}\vect{p^1}=(p,\zeta p_1)$;
\item $\vect{u^P}=M^{P,W}\vect{p^W}=(1-w,w)$.
\end{OneLiners}
Proof of correctness as in \lref[Lemma]{pro:correctness-of-Gcomplementary}.
\end{proof}

The following gadget has multiple output players, denoted by $B_{1},\dots,B_{\beta}$ and representing values $b_1,\dots,b_\beta$. It extracts the first $\beta$ bits of its input, provided the distance of $p_1$ from any multiple of $2^{-\beta}$ is at least $3\beta\epsilon$. 

\begin{lemma}[Linear Bit Extraction Gadget]
\label{pro:correctness-of-Gbit}
For every integer $\beta>0$, there exists a linear bit extraction gadget $G_{\text{bit}}$ of size $O(\beta)$, such that given input $p_1=\sum_{i\in[\beta]}{b_i^*}2^{-i}+\delta$ where $3\beta\epsilon<\delta<2^{-\beta}-3\beta\epsilon$, in every $\epsilon$-well-supported Nash equilibrium where $\epsilon=O\left(2^{-\left(\beta+\log \beta\right)}\right)$, $b_i=b_i^*$ for every $i\in[\beta]$.
\end{lemma}

\begin{proof}
The construction is by combining linear gadgets:\begin{eqnarray*}
x_{1} & = & G_{:=}\left(p_{1}\right)\\
\forall i\mbox{ : }b_{i} & = & G_{>2^{-i}}\left(x_{i}\right)\\
\forall i\mbox{ : }w_{i} & = & G_{*2^{-i}}\left(b_{i}\right)\\
\forall i\mbox{ : }x_{i+1} & = & G_{-}\left(x_{i},w\right)\end{eqnarray*}
The correctness follows from the guarantees of the combined gadgets, and by induction on $i$. See \cite[Lemma 19]{DGP09} for details. 
\end{proof}

\end{document}